\documentclass[]{WileyASNA-v1}

\usepackage[utf8]{inputenc}
\usepackage{calc}
\usepackage{anyfontsize}
\usepackage{hyperref}
\usepackage{endnotes}

\newlength{\okinalen}
\newcommand{\okina}{\hbox to.666\okinalen{\hss`\hss}}

\articletype{Article Type}

\received{2024}
\revised{2025}
\accepted{2025}

\begin{document}

\title{Gaia search for stellar companions of TESS Objects of Interest V}

\author[1]{Markus Mugrauer}

\author[1]{Ann-Kathrin Kollak}

\author[1]{Lara Pietsch}

\author[1]{Kai-Uwe Michel}

\authormark{Mugrauer, Kollak, Pietsch \& Michel}

\address{Astrophysikalisches Institut und Universit\"{a}ts-Sternwarte Jena}

\corres{M. Mugrauer, Astrophysikalisches Institut und Universit\"{a}ts-Sternwarte Jena, Schillerg\"{a}{\ss}chen 2, D-07745 Jena, Germany.\newline \email{markus@astro.uni-jena.de}}

\abstract{In this paper we present the latest results of our ongoing multiplicity survey of (Community) TESS Objects of Interest, using astrometry and photometry from the latest data release of the ESA Gaia mission to detect stellar companions of these stars and to characterize their properties. A total of 92 binary and two hierarchical triple star systems are identified among the 745 target stars whose multiplicity is explored in this study, all at distances of less than 500\,pc around the Sun. As expected for components of gravitationally bound star systems, the targets and the detected companions are at the same distance and share a common proper motion, as shown by their accurate Gaia astrometry. The companions have masses of about 0.12 to 1.6\,$M_\odot$ and are most frequently found in the mass range up to 0.6\,$M_\odot$. The companions have projected separations from the targets between about 110 and 9600\,au. Their frequency is highest and constant from about 300 to 800\,au, decreasing at larger projected separations. In addition to main sequence stars, five white dwarf companions are detected in this study, whose true nature is unveiled by their photometric properties.}


\maketitle

\section{Introduction}

\begin{figure*}
\resizebox{\hsize}{!}{\includegraphics{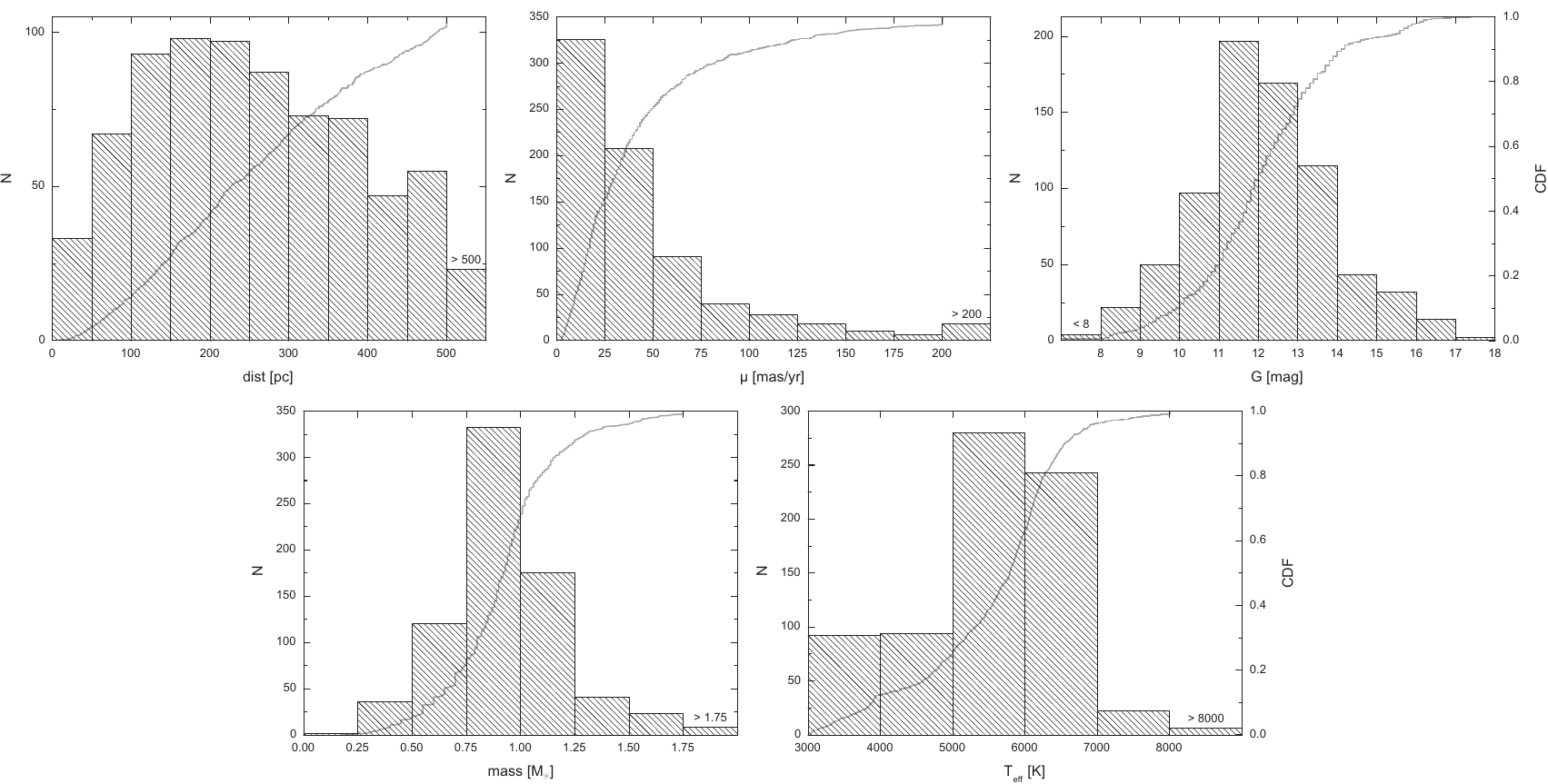}}\caption{The histograms and CDFs of the individual properties of the targets, whose multiplicity is investigated in this study. The histograms and CDFs of the distance $dist$, total proper motion $\mu$, and \mbox{G-band} magnitude are based on the Gaia DR3 data of all 745 targets. Masses and $T_{\rm eff}$ of the targets are taken from the SHC or SHC2 where available, which is the case for 737 targets.}\label{HIST_TARGETS}
\end{figure*}

In the course of our ongoing multiplicity survey of (Community) TESS\footnote{TESS: Transiting Exoplanet Survey Satellite \citep{ricker2015}.} Objects of Interest ((C)TOIs) stellar companions of these stars are detected and their properties are characterized using astrometric and photometric data, originally taken from the 2nd and early 3rd data releases \citep{gaiadr2, gaiaedr3}, and finally from the full 3rd data release \citep[Gaia DR3 from hereon,][]{gaiadr3} of the European Space Agency (ESA) Gaia mission. The first results of the survey have already been published in \cite{mugrauer2020}, \cite{mugrauer2021}, \cite{mugrauer2022}, and \cite{mugrauer2023}. Further information on the survey is summarized in these publications. Since then, many of these (C)TOIs, which were found to be members of multiple star systems in our survey, have been confirmed as exoplanet host stars by follow-up observations, for example: TOI\,128, TOI\,199, TOI\,444, TOI\,470, TOI\,762, TOI\,815, TOI\,837, TOI\,858, TOI\,880, TOI\,907, TOI\,1052, TOI\,1248, TOI\,1450, TOI\,1473, TOI\,1824, TOI\,1855, TOI\,1859, and  TOI\,1937, which are listed in the \verb"Extrasolar Planets Encyclopaedia"\footnote{Online available at: \url{https://exoplanet.eu/}.} \citep[see][and references therein]{schneider2011}. The targets of our survey are all listed in the (C)TOI release of the \verb"Exoplanet" \verb"Follow-up" \verb"Observing" \verb"Program" for TESS\footnote{Online available at: \newline \url{https://exofop.ipac.caltech.edu/tess/view_toi.php} \newline\url{https://exofop.ipac.caltech.edu/tess/view_ctoi.php}.} (ExoFOP-TESS).

In this paper we investigate the multiplicity status of more than 700 (C)TOIs listed in the ExoFOP-TESS, that have not yet been studied in our survey, using data from the Gaia DR3. The following section describes in detail the properties of the selected targets and the search for companions around these stars. Section 3 presents all (C)TOIs with detected companions and discusses the properties of these star systems. Finally, the last section summarizes the current status of our survey and privides an outlook on the project.

\section{Search for stellar companions of (C)TOIs by exploring the Gaia DR3}

In our study, stellar companions of the investigated (C)TOIs are firstly identified as sources that are located at the same distance as the targets and secondly share a common proper motion with these stars. In order to unambiguously detect co-moving companions and confirm their equidistance to the (C)TOIs, we consider in our study only those sources for which a significant measurement of their parallax \mbox{($\pi/\sigma(\pi) > 3$) and proper motion \mbox{($\mu/\sigma(\mu) > 3$)}} is available in the Gaia DR3. Sources with a negative parallax are not taken into account.

In this paper, we explore the multiplicity of 745 (C)TOIs that have not yet been investigated in the course of our survey with Gaia DR3 data, that meet the distance constraint of our survey, described in \cite{mugrauer2019} or \cite{mugrauer2020}, and are therefore selected as targets for this study. Figure\,\ref{HIST_TARGETS}\hspace{-1.5mm} shows the histograms and the cumulative distribution functions (CDFs) of the individual properties of these stars. The distance ($dist$) and total proper motion ($\mu$) of all targets are determined using their precise Gaia DR3 parallax and proper motion in right ascension and declination. The \mbox{G-band} magnitude of all targets is listed in the Gaia DR3, while their mass and effective temperature ($T_{\rm eff}$) are taken from the Starhorse catalog \citep[SHC from here on,][]{anders2019} or from the Starhorse 2 catalog \citep[SHC2 from here on,][]{anders2022} where available, which is the case for 737 stars, that is, the vast majority (\mbox{$\sim99$\,\%}) of all targets. The targets have distances from the Sun between about 7 and 800\,pc, proper motions in the range between about 1 and 840\,mas/yr, \mbox{G-band} magnitudes between 6.3 and 17.4\,mag, masses between about 0.18 and 4.4\,$M_\odot$, and $T_{\rm eff}$ between about 3000 and 12000\,K. According to the CDFs of the individual properties, the targets are most commonly located at distances between about 50 and 400\,pc, have typical proper motions from about 4 to 24\,mas/yr, and \mbox{G-band} magnitudes from about $G=11$ to 13\,mag. The targets are mainly solar-like stars with masses between about 0.8 and 1.2\,$M_\odot$. This population is also recognizable in the \mbox{$T_{\rm eff}$-distribution} of the targets at intermediate temperatures from about 4700 to 6300\,K.\newpage

As described in \cite{mugrauer2019} or \cite{mugrauer2020}, our survey is restricted to companions with projected separations of up to 10000\,au. With $\pi$ the Gaia DR3 parallax of the (C)TOIs this results in an angular search radius for companions around the targets:

\begin{equation}
\mbox{$r [{\rm arcsec}] = 10 \pi[{\rm mas}]$}
\end{equation}
\linebreak
All sources listed in the Gaia DR3 that are located within the search radius used around the targets and have a significant parallax and proper motion are considered as companion-candidates. A total of about 7749 such objects are detected around 745 targets, and their multiplicity is explored in the study, presented here. The companionship of all these candidates is examined using their precise Gaia DR3 astrometry and that of the associated (C)TOIs, following exactly the procedure, which is described in \cite{mugrauer2019} or \cite{mugrauer2020}. The vast majority of these sources can be excluded as companions because they and the associated (C)TOIs do not share a common proper motion and/or are not located at the same distance as these stars. In contrast, 96 candidates can be confidently identified as companions of the (C)TOIs based on their accurate Gaia DR3 astrometry. The properties of these companions and the associated (C)TOIs are described in detail in the next section of this paper.

\begin{figure}
\resizebox{\hsize}{!}{\includegraphics{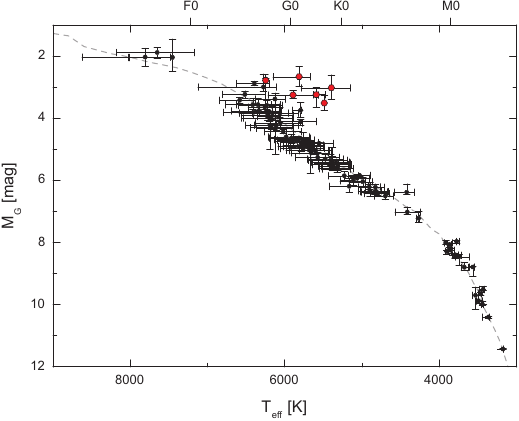}}\caption{The absolute \mbox{G-band} magnitude $M_{\rm G}$ plotted versus the $T_{\rm eff}$ of all (C)TOIs with detected companions, which are presented here. (C)TOIs listed in the SHC or if not available there in the SHC2 with surface gravities \mbox{$\log(g[\rm{cm/s^{-2}}])<4.0$} are shown as red circles, those with larger surface gravities as black circles, respectively. The main sequence from \cite{pecaut2013} is drawn as a grey dashed line for comparison.}\label{HRDCTOIS}
\end{figure}

\section{(C)TOIs and their detected stellar companions}

The mass, $T_{\rm eff}$, and absolute \mbox{G-band} magnitude of all the (C)TOIs with detected companions, presented here, are listed in the SHC or SHC2. We plot these stars in a \mbox{$T_{\rm eff}$-$M_{\rm G}$-diagram} in Figure\,\ref{HRDCTOIS}\hspace{-1.5mm} with the main sequence from \cite{pecaut2013}\footnote{Online available at: \url{https://www.pas.rochester.edu/~emamajek/EEM_dwarf_UBVIJHK_colors_Teff.txt}. The version used here is 2022.04.16.} for comparison. While most targets with detected companions are main sequence stars a few (C)TOIs are (significantly) located above the main sequence. These stars also have a surface gravity of \mbox{$\log(g[\rm{cm/s^{-2}}]) < 4$}, as listed in the SHC or, if not available there, in the SHC2, and are therefore classified as (sub)giant stars.

The parallax, proper motion, apparent \mbox{G-band} magnitude, and extinction estimate $(A_{\rm G})$ of the (C)TOIs and their companions detected in this study are summarized in \mbox{Table\,\ref{TAB_COMP_ASTROPHOTO}\hspace{-1.5mm},}\linebreak which lists a total of 92 binary, and two hierarchical triple star systems.

With the accurate Gaia DR3 astrometry we determine the angular separation ($\rho$) and position angle ($PA$) of all detected companions to the associated (C)TOIs. The derived relative astrometry of the companions is summarized in \mbox{Table\,\ref{TAB_COMP_RELASTRO}\hspace{-1.5mm}}. Its uncertainty remains below about 0.9\,mas in angular separation and 0.02\,$^{\circ}$ in position angle.

Table\,\ref{TAB_COMP_RELASTRO}\hspace{-1.5mm} also lists the parallax difference $\Delta \pi$ between the (C)TOIs and their companions, together with its significance $sig\text{-}\Delta\pi$, which is also calculated taking into account the astrometric excess noise of each object. The same table contains for each companion its differential proper motion $\mu_{\rm rel}$ relative to the associated (C)TOI with its significance, and its common proper motion ($cpm$) index. The $cpm$-$index$, as defined in \cite{mugrauer2019} or \cite{mugrauer2020}, characterizes the degree of common proper motion of a detected companion with the associated (C)TOI:

\begin{equation}
cpm\text{-}index = | \overrightarrow{\mu}_{(C)TOI} + \overrightarrow{\mu}_{Comp}| / \mu_{rel}
\end{equation}
\linebreak
with $\overrightarrow{\mu}_{(C)TOI}$ the proper motion of the (C)TOI, and $\overrightarrow{\mu}_{Comp}$ the proper motion of the companion, as well as its differential proper motion $\mu_{rel}=| \overrightarrow{\mu}_{(C)TOI} - \overrightarrow{\mu}_{Comp}|$.

The parallaxes of the individual components of the star systems, detected in this study, do not differ significantly from each other \mbox{($sig\text{-}\Delta\pi < 3$)} when the astrometric excess noise is considered. This clearly proves the equidistance of the detected companions and the (C)TOIs, as expected for components of physically associated star systems. All of the detected companions have a \mbox{$cpm\text{-}index \geq 10$}, that is, the detected companions and the associated (C)TOIs clearly form common proper motion pairs, as expected for gravitationally bound star systems.

\begin{figure}
\resizebox{\hsize}{!}{\includegraphics{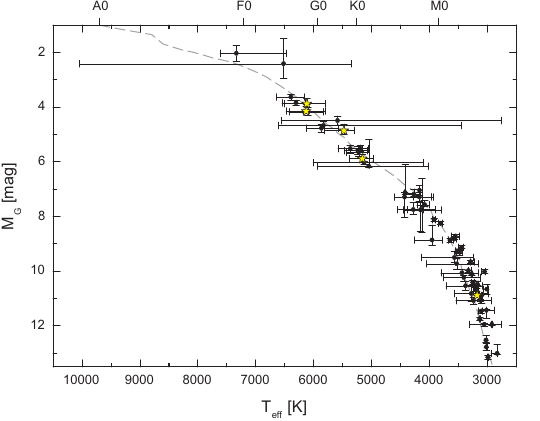}}\caption{The absolute \mbox{G-band} magnitude $M_{\rm G}$ plotted over the $T_{\rm eff}$ of all detected companions whose $T_{\rm eff}$ are listed in either the SHC or SHC2, or for which Apsis GSP-Phot Aeneas temperature estimates are available. Companions, which are the primary components in their star system, are shown as yellow star symbols. The main sequence from \cite{pecaut2013} is drawn as a dashed grey line for comparison.}\label{HRDCOMPS}
\end{figure}

The absolute \mbox{G-band} magnitude of all detected companions is taken from the SHC or, if not available there, from the SHC2, indicated by the \texttt{SHC2} flag in \mbox{Table\,\ref{TAB_COMP_PROPS}\hspace{-1.5mm}}. Where the absolute magnitude of the companions is not listed in these catalogs, it is inferred from their apparent \mbox{G-band} photometry and the parallax of the (C)TOIs, and the SHC (if available) or the SHC2 \mbox{G-band} extinction estimate, otherwise the $A_{\rm G}$ estimate from Vizier. The extinction estimate of the companions is used if available, otherwise that of the (C)TOIs, as indicated in \mbox{Table\,\ref{TAB_COMP_ASTROPHOTO}\hspace{-1.5mm}}. In general, the used $A_{\rm G}$ estimates are in good agreement with those determined by other authors, as listed in the \verb"VizieR" database, or those derived from dust infrared emission maps \citep[see e.g.,][]{schlegel1998, schlafly2011}. The deviation of the individual $A_{\rm G}$ estimates is on average 0.16\,mag, which, as expected, matches the mean uncertainty of the used extinction estimates. In this context, it should be emphasized that the derived absolute magnitude of the companions and thus also their mass and $T_{\rm eff}$ also depend on the used extinction estimate. If this estimate is set too high/low, the derived mass and $T_{\rm eff}$ of the companions are also higher/lower. For the typical companions detected in this study (average mass of about 0.5\,$M_\odot$), a deviation in $A_{\rm G}$ of 0.16\,mag leads to a mass shift of about 0.03\,$M_\odot$ and about 50\,K in $T_{\rm eff}$, which is smaller than the mean uncertainty of both quantities (about 0.05\,$M_\odot$, and 200\,K), as listed in \mbox{Table\,\ref{TAB_COMP_PROPS}\hspace{-1.5mm}}.

The projected separation of all companions is determined from their angular separation from the associated (C)TOI and the parallax of these stars.

The mass and $T_{\rm eff}$ of the companions presented here, including their uncertainties, are from the SHC or the SHC2 (indicated by the \texttt{SHC2} flag in \mbox{Table\,\ref{TAB_COMP_PROPS}\hspace{-1.5mm}}) where available, which is the case for about 63\,\% of all detected companions. We plot these companions in Figure\,\ref{HRDCOMPS}\hspace{-1.5mm} in a \mbox{$T_{\rm eff}$-$M_{\rm G}$-diagram} together with those companions for which an estimate of their $T_{\rm eff}$ is determined by Apsis GSP-Phot\footnote{The General Stellar Parametrizer from Photometry (GSP-Phot) is one of the major component modules in the astrophysical parameters inference system (Apsis) of Gaia.} using the MCMC algorithm Aeneas\footnote{GSP-Phot Aeneas estimates the $T_{\rm eff}$ of detected sources from their low-resolution BP/RP spectra, apparent G-band magnitude and parallax.}, as indicated by the $\texttt{AEN}$ flag in \mbox{Table\,\ref{TAB_COMP_PROPS}\hspace{-1.5mm}}. As illustrated in Figure\,\ref{HRDCOMPS}\hspace{-1.5mm}, the absolute photometry and $T_{\rm eff}$ of all these companions are in good agreement with those expected for main sequence stars.

\begin{figure*}
\includegraphics[width=\textwidth]{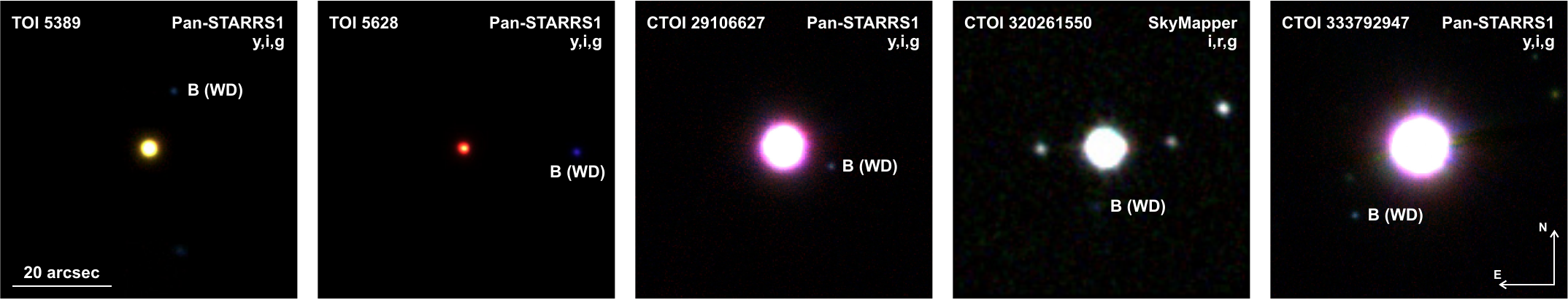}
\caption{(RGB)-color images of TOI\,5389, TOI\,5628, CTOI\,29106627, and CTOI\,333792947 with their white dwarf companions composed of y-, i- and \mbox{g-band} Pan-STARRS images. The image of CTOI\,320261550 and its white dwarf companion is made of i-, r-, and g-band images taken in the SkyMapper Southern Sky Survey.}\label{PICS}
\end{figure*}

For the remaining 36 companions their mass and $T_{\rm eff}$ are determined from their absolute \mbox{G-band} magnitude via interpolation ($\texttt{inter}$ flag in Table\,\ref{TAB_COMP_PROPS}\hspace{-1.5mm}) using the \mbox{$M_{\rm G}$-mass}- and \mbox{$M_{\rm G}$-$T_{\rm eff}$-relation} from \cite{pecaut2013}, assuming that these companions are main sequence stars. In order to test this hypothesis, we compare the obtained $T_{\rm eff}$ of the companions either with their Apsis GSP-Phot Aeneas temperature estimate, if available, or with the $T_{\rm eff}$ of the companions obtained from their \mbox{$(B_{\rm P}-R_{\rm P})$} color and reddening estimate \mbox{$E(B_{\rm P} - R_{\rm P})$}\footnote{The reddening of an object is estimated from its extinction $A_{\rm G}$, using the relation \mbox{$A_{\rm G} / E(B_{\rm P} - R_{\rm P}) = 1.890 \pm 0.015$} from \cite{wang2019}.}, using the \mbox{$(B_{\rm P}-R_{\rm P})_0$-$T_{\rm eff}$-relation} from \cite{pecaut2013}.

For all but five of these companions, their $T_{\rm eff}$, determined from their absolute G-band magnitude and assuming that they are main sequence stars, agrees well with their Apsis GSP-Phot Aeneas temperature estimate or with the temperature derived from their color. The typical deviation of the different temperature estimates is about 390\,K, which is consistent with the precision of the derived $T_{\rm eff}$. We therefore conclude that all of these companions are main sequence stars.

In addition, we also compare the Gaia DR3 \mbox{$(B_{\rm P}-R_{\rm P})$} color of the (C)TOIs and their companions (if any), indicated by the $\texttt{BPRP}$ flag in \mbox{Table\,\ref{TAB_COMP_PROPS}\hspace{-1.5mm}}. For main sequence stars, companions fainter/brighter than the (C)TOIs are expected to appear redder/bluer than the stars, and this is true for most of the discovered companions, with the exception of TOI\,5389\,B, TOI\,5628\,B, CTOI\,29106627\,B, and CTOI\,320261550\,B. Three of these four companions were also observed with the Panoramic Survey Telescope and Rapid Response System (Pan-STARRS) and their color composite images are shown in Figure\,\ref{PICS}\hspace{-1.5mm}. In these images, the faint companions are clearly visible as bluish sources next to the associated much brighter (C)TOIs. The same applies to the faint companion CTOI\,320261550\,B, imaged as part of the SkyMapper Southern Sky Survey. The color composite image of this companion is also shown in Figure\,\ref{PICS}\hspace{-1.5mm}.

The photometric properties of all these companions are summarized in \mbox{Table\,\ref{TAB_WDS_PROPS}\hspace{-1.5mm}}. The companions are several magnitudes fainter than the (C)TOIs, but appear bluer than these stars. The temperatures of the companions derived from their colors are significantly higher (by about 3600 to 8600 K) than the temperatures, derived from their absolute magnitudes in the G-band, assuming that they are main sequence stars.

\begin{table*}[h!]
\caption{In this table we summarize the photometric properties of all the white dwarf companions, detected in this study. For each companion we list the color difference $\Delta(B_{\rm P}-R_{\rm P})$ and the \mbox{G-band} magnitude difference $\Delta G$ to the associated (C)TOI, its apparent $(B_{\rm P}-R_{\rm P})$ color, as well as its derived intrinsic color $(B_{\rm P}-R_{\rm P})_{0}$ and $T_{\rm eff}$.}\label{TAB_WDS_PROPS}
\begin{center}
\begin{tabular}{lccccc}
\hline
Companion         & $\Delta G$        & $\Delta (B_{\rm P}-R_{\rm P})$ & $(B_{\rm P}-R_{\rm P})$ & $(B_{\rm P}-R_{\rm P})_0$      & $T_{\rm eff}$\\
                  & [mag]             & [mag]                          & [mag]                   & [mag]                          & [K]\\
\hline
TOI\,5389\,B      & $4.266 \pm 0.007$ & $-1.718 \pm 0.118$             & $0.622 \pm 0.118$       & $0.620_{-0.128}^{+0.118}$      & $6425_{-399}^{+389}$\\
TOI\,5628\,B      & $1.625 \pm 0.004$ & $-2.582 \pm 0.014$             & $0.079 \pm 0.011$       & $0.042_{-0.016}^{+0.022}$      & $9008_{-173}^{+131}$\\
CTOI\,29106627\,B & $8.137 \pm 0.007$ & $-0.551 \pm 0.203$             & $0.559 \pm 0.203$       & $0.513_{-0.213}^{+0.208}$      & $6764_{-716}^{+888}$\\
CTOI\,320261550\,B& $7.223 \pm 0.004$ & $-0.926 \pm 0.062$             & $-0.117 \pm 0.062$~~\,  & $-0.245_{-0.113}^{+0.088}$~~\, & $11841_{-797}^{+1031}$\\
CTOI\,333792947\,B& $9.578 \pm 0.007$ & $~~\,0.229 \pm 0.205$          & $1.052 \pm 0.205$       & $1.017_{-0.217}^{+0.208}$      & $5165_{-361}^{+708}$\\
\hline
\end{tabular}
\end{center}
\label{table_WD}
\end{table*}

\begin{figure}
\includegraphics[width=\linewidth]{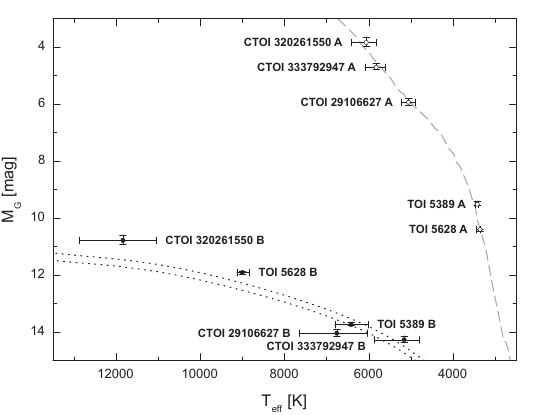}\caption{The absolute magnitude $M_{\rm G}$ plotted against the $T_{\rm eff}$ of the components of all star systems with detected white dwarf companions. The grey dashed line shows the main sequence, the black dotted lines the evolutionary mass tracks of DA white dwarfs with masses of 0.5 and 0.6\,$M_\odot$. The primaries of the systems are shown as white circles, the white dwarf secondaries as black circles, respectively.}\label{HRD_WDS}
\end{figure}

These companions, along with the other components of their star system, are plotted in a $M_{\rm G}$-$T_{\rm eff}$-diagram in \mbox{Figure\,\ref{HRD_WDS}\hspace{-1.5mm}}. The main sequence from \cite{pecaut2013} and the mass tracks from the Bergeron et al. evolutionary models\footnote{The models are available online at: \url{https://www.astro.umontreal.ca/~bergeron/CoolingModels/}. The version used here is 2021.01.13. For more details about the models, see: \citealp{bedard2020}; \citealp{bergeron2011}; \citealp{blouin2018}; \citealp{holberg2006}; \citealp{kowalski2006}; \citealp{tremblay2011}.} are plotted in this diagram for comparison. While the brighter primary components of these systems are main sequence stars, the faint secondary stars are all located well below the main sequence, and their Gaia photometry is the most consistent with that expected for white dwarfs. We therefore conclude that these companions are white dwarfs, as indicated by the $\texttt{WD}$ flag in  \mbox{Table\,\ref{TAB_COMP_PROPS}\hspace{-1.5mm}}.

Another white dwarf, CTOI\,333792947\,B, which is also shown in Figure\,\ref{HRD_WDS}\hspace{-1.5mm}, is revealed in this study because of its comparable color ($\Delta (B_{\rm P}-R_{\rm P}) = 0.229 \pm 0.205$) but large magnitude difference to its primary star ($\Delta G \sim 9.6$\,mag).

\begin{figure*} [h]
\includegraphics[width=\linewidth]{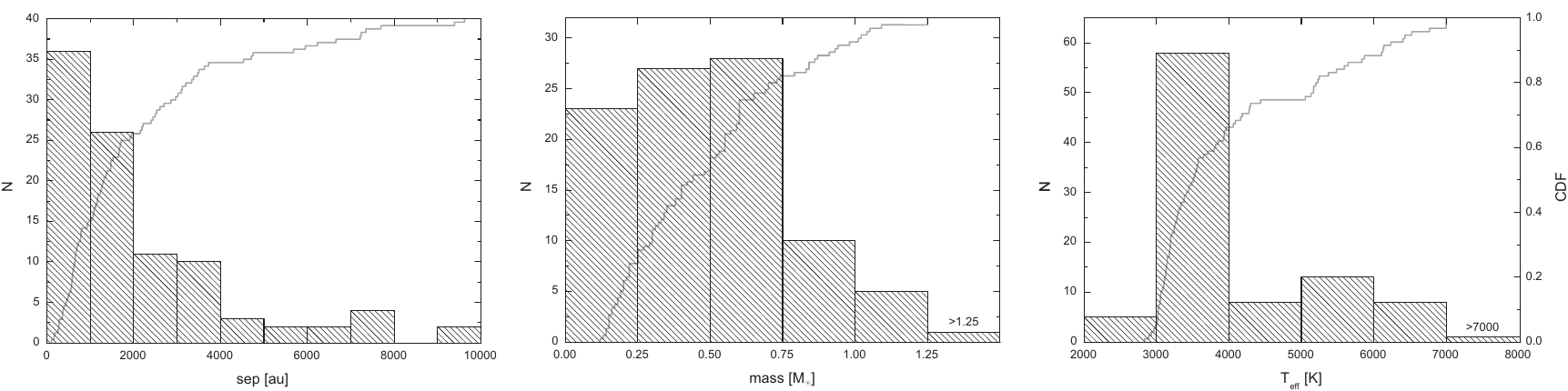}\caption{The histograms and CDFs of the projected separation, mass, and $T_{\rm eff}$ of all detected companions, presented here.}\label{HIST_COMPS}
\end{figure*}

Figure\,\ref{HIST_COMPS}\hspace{-1.5mm} shows the histograms and CDFs of the properties of all the companions detected in this study. The companions have angular separations from the (C)TOIs ranging from about 0.8 to 111\,arcsec, corresponding to projected separations of 113 to 9611\,au. According to the underlying CDF, the frequency of companions is highest and constant between about 300 and 800\,au, and decreases at larger projected separations. Half of all companions have projected separations of less than 1300\,au. A total of three stellar systems with projected separations below 200\,au are detected, namely: TOI\,5319\,AB, CTOI\,199572211\,AB, CTOI\,287643871\,AB, that is, these systems are the most challenging environments for planet formation identified in this study.

\begin{figure}[h]
\includegraphics[width=\linewidth]{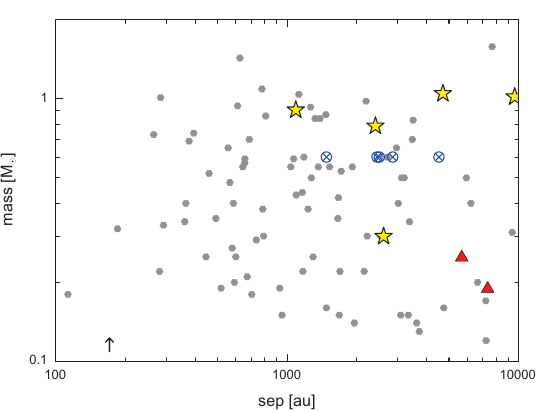}\caption{The mass of all detected companions, presented here, plotted over their projected separation. Companions that are the primary component of their star system are shown as yellow star symbols, secondaries as grey circles, and tertiary components as red triangles, respectively. Detected white dwarf companions, for which a mass of 0.6\,$M_\odot$ is assumed, are plotted as blue crossed circles. Note that the symbols of the white dwarf companions TOI\,5389\,B, and TOI\,5628\,B overlap at a separation of about 2500\,au. The separation of the companion CTOI\,287643871\,B, for which no mass could be determined because its G-band magnitude is not listed in the Gaia DR3, is marked by a black arrow.}\label{SEPMASS}
\end{figure}

The companion masses range from about 0.12 to 1.6\,$M_\odot$ (average mass: \mbox{$\sim 0.5\,M_\odot$}). The highest companion frequency in the CDF is found in the mass range up to 0.6\,$M_\odot$, corresponding to mid-M to late-K dwarfs, according to the relation between mass and spectral type (SpT) from \cite{pecaut2013}. At higher masses, the companion frequency decreases continuously towards higher masses. This peak in the companion population is also present in the distribution of their $T_{\rm eff}$, where the companion frequency is highest at temperatures between about 3000 and 4000 K. As can be seen from the \mbox{separation-mass diagram} in Figure\,\ref{SEPMASS}\hspace{-1.5mm}, of the 96 companions presented here, five are the primary, 89 the secondary and two the tertiary components of their star system.

\begin{figure}
\includegraphics[width=\linewidth]{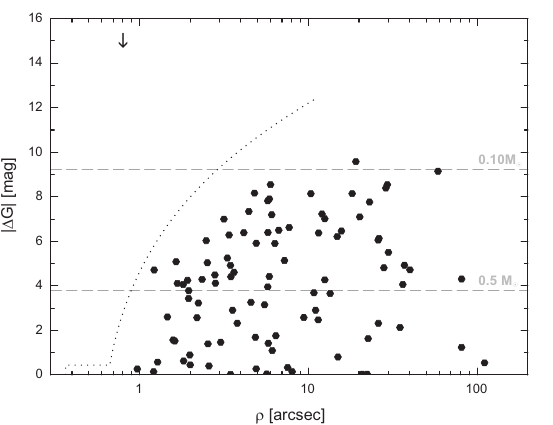}\caption{The \mbox{G-band} magnitude difference $|\Delta G|$ of all detected companions, presented here, plotted versus their angular separation $\rho$ from the associated (C)TOIs. The Gaia detection limit, determined by \cite{mugrauer2023}, is drawn as a dotted line for comparison. The grey dashed horizontal lines show the expected average magnitude differences for companions of 0.1 and 0.5\,$M_\odot$, respectively. The angular separation of the companion CTOI\,287643871\,B, for which no G-band magnitude is listed in the Gaia DR3, is indicated by a black arrow.}\label{LIMIT}
\end{figure}

\begin{figure*} [h!]
\begin{center}\includegraphics[width=\textwidth]{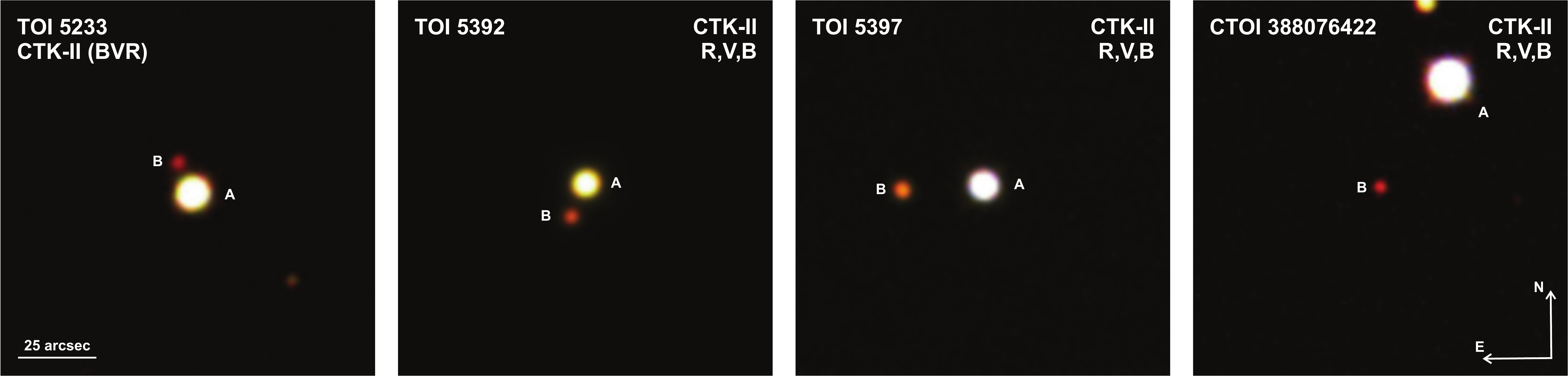}\end{center}\caption{(RGB)-color images of the binary systems TOI\,5233\,AB, TOI\,5392\,AB, TOI\,5397\,AB, and CTOI\,388076422\,BA, composed of R-, V-, and \mbox{B-band} images, taken with the CTK-II at the University Observatory Jena.}\label{PIC_CTKII}
\end{figure*}

In order to characterize the detection limit, achieved in this study, we plot the magnitude difference of all detected companions over their angular separation from the associated (C)TOIs, as shown in Figure\,\ref{LIMIT}\hspace{-1.5mm}. For comparison, we show the Gaia detection limit, determined by \cite{mugrauer2023}. As expected, all detected companions, presented here, remain within this limit.

The magnitude difference expected between the targets of this study and low-mass main sequence companions (shown as grey dashed lines in Figure\,\ref{LIMIT}\hspace{-1.5mm}) is estimated using the expected absolute \mbox{G-band} magnitude of these stars, as listed in \cite{pecaut2013}, and the average of the absolute \mbox{G-band} magnitude of our targets (\mbox{$M_{\rm G} \sim 5.0$\,mag}). As can be seen in Figure\,\ref{LIMIT}\hspace{-1.5mm}, a magnitude difference of about 4\,mag is achieved at an angular separation of about 0.9\,arcsec around the targets of this study. This allows the detection of companions with masses down to about 0.5\,$M_\odot$ (the average mass of all companions detected) separated from the (C)TOIs by more than 230\,au. In addition, all stellar companions with masses above 0.1\,$M_\odot$ are detectable beyond $\sim 3$\,arcsec, which corresponds to a projected separation of about 770\,au at the average target distance of 255\,pc.

\section{Summary and Outlook}

Several ground-based either Seeing limited or high contrast imaging surveys have been performed in the past to study the multiplicity of exoplanet host stars, among others, for example, \cite{mugrauer2014}, or \cite{mugrauer2015}, and \cite{ginski2016}. Here we present the latest results of our survey using data from ESA's Gaia mission, which is based on methods first applied on Gaia data in the multiplicity survey of exoplanet host stars, performed by \cite{mugrauer2019}. Later, other comprehensive stellar multiplicity surveys were also carried out using Gaia data \citep[see e.g.][]{elbadry2021, kervella2022}.

In the present study, we use Gaia DR3 data to search for stellar companions of 745 (C)TOIs announced in the (C)TOI release of the ExoFOP-TESS and whose multiplicity has not yet been explored in the course of our survey. A total of 7749 sources were detected around 567 targets with accurate astrometric solutions in the Gaia DR3, while no companion-candidates were found around the remaining 178 targets of this study within the applied search radius. In total, new co-moving companions were detected around 94 of all targets whose multiplicity is studied here. In addition, companions around another 20 (C)TOIs were found in the Gaia DR3, but were already identified in the Gaia DR2 by \cite{mugrauer2019} and \cite{mugrauer2020} or most recently by \cite{michel2024}, who used Gaia DR3 data to study the multiplicity of confirmed exoplanet host stars. The multiplicity rate of the examined (C)TOIs is thus at least $15.3 \pm1.4$\,\%, which agrees well with the (C)TOI multiplicity rate of $14.9 \pm1.1$\,\%, recently reported by \cite{mugrauer2023}.

Color-composite images of some of the detected star systems are shown in Figure\,\ref{PIC_CTKII}\hspace{-1.5mm}, taken with the Cassegrain-Teleskop-Kamera-II \citep[CTK-II from hereon,][]{mugrauer2016} at the University Observatory Jena. In addition to 92 double stars, two hierarchical triple star systems are detected in which the (C)TOIs have both a close and a distant stellar companion. In addition, an astrometric non-single star solution for the co-moving companion TOI\,4660\,B is listed in the Gaia DR3 (marked with the flag \texttt{NSS} in \mbox{Table\,\ref{TAB_COMP_PROPS}\hspace{-1.5mm}}). Based on 230 astrometric measurements made with Gaia for this star, a significant (\mbox{$sig>30$}) time-varying acceleration ($\dot{\mu} = 2.75 \pm 0.13\,\text{mas/yr}^2$ and $\ddot{\mu} = 17.85 \pm 0.51\,\text{mas/yr}^3$) of the companion in the plane of the sky is detected, indicating that it is itself a close binary system. The orbital period of this system must be significantly longer than the 34-month period on which the Gaia DR3 is based. We therefore classify TOI\,4660 as a potential hierarchical triple star system, whose triple nature needs to be confirmed by follow-up observations.

The Washington Double Star catalog \citep[WDS from hereon, ][]{wds}, which is available in the \verb"VizieR" database, list 18 of the companions identified in this study either as co-moving companions or as companion-candidates of the (C)TOIs, so confirmation of their companionship is required, which is finally provided by this study. Although the WDS is currently the most complete catalogue of multiple star systems available, containing relative astrometric measurements of multiple star systems over a period of more than 300 years, 78 (i.e., about 81\,\% of all) companions, detected in this study, are not listed in the WDS and marked with the $\star$ flag in the \mbox{Table\,\ref{TAB_COMP_RELASTRO}\hspace{-1.5mm}}. Additional companions that are not included in the WDS because they were detected in the Gaia data releases are listed in the WDS Supplemental Catalogue (WDSS), which is not included in the \verb"VizieR" database but is available for download in its latest version\footnote{Online available at: \url{https://www.astro.gsu.edu/wds/Supplement/wdss_summ.txt}.}. Companions detected in this survey that are listed neither in the WDS nor WDSS are indicated with the $\star\star$ flag in \mbox{Table\,\ref{TAB_COMP_RELASTRO}\hspace{-1.5mm}}. This shows the great potential of the ESA Gaia mission for stellar multiplicity surveys, especially for the detection of stellar companions on wide orbits, as shown by the derived detection limit of this study in Figure\,\ref{LIMIT}\hspace{-1.5mm}. On average, all stellar companions with masses above about 0.1\,$M_\odot$ are detectable around the targets of this study at angular separations larger than about 3\,arcsec (or 770\,au of projected separation), and about 68\,\% of all detected companions have such separations. At the average distance of our targets of 255\,pc, this mass limit also corresponds to the faintest detectable companions ($G=21$\,mag), which exhibit significant measurements of parallax \mbox{($\pi/\sigma(\pi) > 3$) and proper motion \mbox{($\mu/\sigma(\mu) > 3$)}} in the Gaia DR3. Overall, companions are detected at projected separations between about 110 and 9600\,au, and the companion frequency is highest at separations between about 300 and 800\,au, decreasing significantly at larger projected separations.

\begin{figure}
\begin{center}\includegraphics[width=\linewidth]{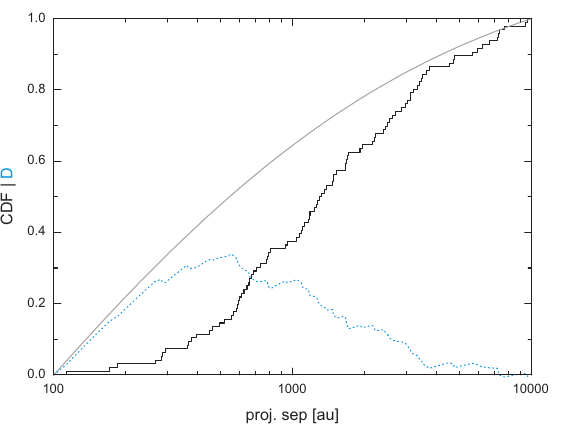}\end{center}\caption{The CDF of the projected separation of all companions, detected in this study (black line), and that of stellar companions of solar type stars (grey line) for the range of projected separation between 100 and 10000\,au. The discrepancy between both CDFs is shown as blue line. The increasing discrepancy between both CDFs within the first few hundred au, illustrates the lack of close companions found in this study, based on the Gaia direct imaging detection performance.}\label{PIC_CDFs}
\end{figure}

The apparent lack of close companions with projected separations below 300\,au is due to the Gaia detection limit, as illustrated in Figure\,\ref{LIMIT}\hspace{-1.5mm}, according to which the direct detection of low-mass stellar companions in particular is only possible with Gaia at projected separations beyond a few hundred au. This also becomes clear when comparing the CDF of the projected separation of all companions detected in this study with that of companions of solar like stars (the typical targets of our survey) as determined by \cite{raghavan2010}, both of which are shown in \mbox{Figure\,\ref{PIC_CDFs}\hspace{-1.5mm}.}

Within the projected separation range considered here (100 - 10000\,au), the CDFs are significantly different from each other ($D_{max}=0.338$ at $\sim550$\,au $\rightarrow$ KS-Test: $\alpha<10^{-9}$). The lack of close companions is evident in an increasing discrepancy of the CDFs within the first few hundred au. Beyond about 600\,au, the discrepancy decreases in agreement with the Gaia detection limit, which allows the detection of even low-mass companions down to $0.1\,M_{\odot}$ in this range of projected separation. This highlights the need of high contrast adaptive optics and lucky-imaging surveys to detect missing close companions of the targets, such as those carried out, for example, by \cite{lillobox2024}, or \cite{matson2025} specifically for TOIs, or for exoplanet host stars in general, for example, by \cite{ginski2021}, or \cite{schlagenhauf2024}.

The companions found in this study have masses of about 0.12 to 1.6\,$M_\odot$ and are most frequent in the mass range up to 0.6\,$M_\odot$. In addition to low-mass main sequence stars (mainly mid-M to late-K dwarfs), five white dwarfs are identified in this study as companions of (C)TOIs, whose true nature is unveiled by their photometric properties.

For 72 (i.e., 75\,\% of all) companions presented here, a significant \mbox{($sig\text{-}\mu_{\rm rel} \geq 3$)} differential proper motion $\mu_{\rm rel}$ relative to the associated (C)TOIs is detected. Using the approximation described in \cite{mugrauer2019}, we derive the escape velocity $\mu_{\rm esc}$ of all these companions. The differential proper motion of the majority of these companions is in agreement with orbital motion. In contrast, the differential proper motion of 9 companions significantly exceeds the expected escape velocity. Because these companions all have a high degree of common proper motion ($cpm\text{-}index\geq12$), this could indicate a higher degree of multiplicity as described in \cite{mugrauer2019}. Follow-up (high contrast imaging) observations are required to further investigate the multiplicity status of all these particular systems and their companions, which are summarized in \mbox{Table\,\ref{table_triples}\hspace{-1.5mm}}.

\begin{table}[h] \caption{In this table we list all detected companions (sorted by their identifier) whose differential proper motion $\mu_{\rm rel}$ relative to the associated (C)TOIs significantly exceeds the expected escape velocity~$\mu_{\rm esc}$.}
\begin{center}
\begin{tabular}{lccc}
\hline
Companion          & $\mu_{\rm rel}$  & $\mu_{\rm esc}$ &\\
                   & [mas/yr]         & [mas/yr] &\\
\hline
TOI\,5122\,B       & $6.43 \pm 0.22 $ & $2.55 \pm 0.01$ &\\
TOI\,5285\,A       & $3.85 \pm 0.23 $ & $1.82 \pm 0.01$ &\\
TOI\,5540\,B       & $2.17 \pm 0.05 $ & $1.93 \pm 0.02$ &\\
TOI\,5544\,B       & $5.48 \pm 0.20 $ & $4.85 \pm 0.05$ &\\
TOI\,5606\,A       & $1.54 \pm 0.02 $ & $0.70 \pm 0.01$ &\\
CTOI\,51099561\,B  & $4.82 \pm 0.06 $ & $4.37 \pm 0.02$ &\\
CTOI\,246976997\,B & $6.28 \pm 0.05 $ & $3.08 \pm 0.02$ &\\
CTOI\,257554718\,B & $2.31 \pm 0.29 $ & $0.99 \pm 0.01$ &\\
CTOI\,376973804\,B & $12.43 \pm 0.09\,\,\,$ & $7.34 \pm 0.05$ &\\
\hline
\end{tabular}
\end{center}
\label{table_triples}
\end{table}

The survey, the latest results of which are presented in this paper, is an ongoing project whose target list is continuously growing due to the increasing number of potential exoplanet host stars identified by the TESS mission. In the course of our survey, the multiplicity status of several thousand (C)TOIs has already been investigated, and all these targets are listed in \mbox{Table\,S1}, which is available online as Supporting Information Material to this article. The properties of the detected stellar companions of the targets of this survey are regularly reported in this journal, online in the \verb"VizieR" database, and on the website of the survey\footnote{Online available at: \url{https://www.astro.uni-jena.de/Users/markus/Multiplicity_of_(C)TOIs.html}.}. The results of our survey, in combination with those of ongoing high-contrast imaging observations of (C)TOIs, which can detect close companions at projected separations down to only a few au, will complete our knowledge of the multiplicity of all these potential exoplanet host stars.

\section{Supporting information}

Table S1: This tab-separated table contains for all targets, whose multiplicity has been studied in the course of this survey so far, their TIC identifier, and J2000 equatorial coordinates, as listed in the TESS Input Catalog  \citep[TIC version 8.2, ][]{paegert2021}, which is online available in the \verb"VizieR" database.

\bibliography{mugrauer}

\section*{Acknowledgments}

We use data from:

(1) the ESA Gaia mission (\url{https://www.cosmos.esa.int/gaia}), processed by the Gaia Data Processing and Analysis Consortium (DPAC, \url{https://www.cosmos.esa.int/web/gaia/dpac/consortium}). The DPAC is funded by national institutions, in particular those participating in the Gaia Multilateral Agreement.

(2) the \verb"Exoplanet Follow-up Observing Program" website, operated by the California Institute of Technology, on behalf of the National Aeronautics and Space Administration, under the Exoplanet Exploration Program.

(3) the \verb"Simbad" and \verb"VizieR" databases operated at the CDS in Strasbourg, France.

(4) the \verb"Extrasolar Planets Encyclopaedia".

(5) the Pan-STARRS1 surveys, made possible by contributions from the Institute for Astronomy, the University of Hawaii, the Pan-STARRS Project Office, the Max-Planck Society and its participating institutes, the Max Planck Institute for Astronomy, Heidelberg and the Max Planck Institute for Extraterrestrial Physics, Garching, The Johns Hopkins University, Durham University, the University of Edinburgh, the Queen's University Belfast, the Harvard-Smithsonian Center for Astrophysics, the Las Cumbres Observatory Global Telescope Network Incorporated, the National Central University of Taiwan, the Space Telescope Science Institute, and the National Aeronautics and Space Administration under Grant No. NNX08AR22G issued through the Planetary Science Division of the NASA Science Mission Directorate, the National Science Foundation Grant No. AST-1238877, the University of Maryland, Eotvos Lorand University (ELTE), and the Los Alamos National Laboratory. The Pan-STARRS1 Surveys are archived at the Space Telescope Science Institute (STScI) and are available through MAST, the Mikulski Archive for Space Telescopes. Additional support for the Pan-STARRS1 public science archive is provided by the Gordon and Betty Moore Foundation.

(6) the SkyMapper Southern Sky Survey, whose national facility capability has been funded through ARC LIEF grant LE130100104 from the Australian Research Council, awarded to the University of Sydney, the Australian National University, Swinburne University of Technology, the University of Queensland, the University of Western Australia, the University of Melbourne, Curtin University of Technology, Monash University and the Australian Astronomical Observatory. SkyMapper is owned and operated by The Australian National University's Research School of Astronomy and Astrophysics. The survey data were processed and provided by the SkyMapper Team at ANU. The SkyMapper node of the All-Sky Virtual Observatory (ASVO) is hosted at the National Computational Infrastructure (NCI). Development and support of the SkyMapper node of the ASVO has been funded in part by Astronomy Australia Limited (AAL) and the Australian Government through the Commonwealth's Education Investment Fund (EIF) and National Collaborative Research Infrastructure Strategy (NCRIS), particularly the National eResearch Collaboration Tools and Resources (NeCTAR) and the Australian National Data Service Projects (ANDS).

(7) the University Observatory Jena, which is operated by the Astrophysical Institute of the Friedrich-Schiller-University.

\setcounter{table}{2}

\begin{table*}[h]
\caption{In this table we summarize for all (C)TOIs (listed first, including their TIC identifier) and their detected co-moving companions their Gaia DR3 parallax $\pi$, proper motion $\mu$ in right ascension and declination, astrometric excess noise $epsi$, G-band magnitude, as well as the used G-Band extinction estimate $A_{\rm G}$ from the SHC or if not available the G-Band extinction estimate, as listed either in the SHC2 or in the VizieR database, indicated with \texttt{SHC2}, and \texttt{\maltese}, respectively.}\label{TAB_COMP_ASTROPHOTO}
\resizebox{\hsize}{!}{\begin{tabular}{ccccccccc}
\hline
TOI       & TIC          & $\pi$  & $\mu_\alpha\,cos(\delta)$&$\mu_\delta$& $epsi$& $G$& $A_{\rm G}$& \\
          &              & [mas]  & [mas/yr] & [mas/yr] & [mas] & [mag] & [mag] & \\
\hline
4580\,A   & 219742885    & $14.7820\pm0.0092\,\,\,$ & $93.028\pm0.012$   & $-259.051\pm0.013\,\,\,\,\,\,$ & $0.044$ & $\,\,\,9.0650\pm0.0028$  &                              &\\
4580\,B   &              & $14.7760\pm0.0749\,\,\,$ & $91.284\pm0.096$   & $-257.824\pm0.104\,\,\,\,\,\,$ & $0.434$ & $17.4529\pm0.0029$ & $0.3114_{-0.0277}^{+0.3250}$ &\\
\hline
4601\,A   & 349071261    & $4.5498\pm0.0184$  & $\,\,\,4.475\pm0.020$    & $-14.914\pm0.015\,\,\,$  & $0.087$ & $10.7645\pm0.0028$ &                              &\\
4601\,B	  &              & $4.5474\pm0.0265$  & $\,\,\,4.333\pm0.032$    & $-14.730\pm0.022\,\,\,$  & $0.015$ & $14.4172\pm0.0028$ & $0.1233_{-0.0555}^{+1.0486}$ &\\
	\hline
4609\,A   & 118820486    & $2.9102\pm0.0218$  & $-0.462\pm0.026\,$   & $-6.034\pm0.017$   & $0.109$ & $10.8653\pm0.0028$ &                              &\\
4609\,B	  &              & $2.8447\pm0.0378$  & $-0.082\pm0.047\,$   & $-6.275\pm0.031$   & $0.232$ & $10.8968\pm0.0028$ & $1.1640_{-0.2847}^{+0.2851}$ &\\
	\hline
4642\,B   & 336961891    & $42.4443\pm0.0196\,\,\,$ & $\,92.003\pm0.023$   & $-122.038\pm0.018\,\,\,\,\,\,$ & $0.092$ & $13.2717\pm0.0028$ &                              &\\
4642\,A   &              & $42.5314\pm0.0314\,\,\,$ & $\,94.018\pm0.035$   & $-120.392\pm0.026\,\,\,\,\,\,$ & $0.220$ & $12.7336\pm0.0029$ & $0.0004_{-0.0004}^{+0.0154}$ &\texttt{\maltese}\\
	\hline
4656\,A   & 120523488    & $2.3036\pm0.0231$  & $\,\,\,1.995\pm0.017$    & $-8.516\pm0.018$   & $0.083$ & $13.9181\pm0.0028$ & $0.3610_{-0.1480}^{+0.1785}$ &\\
4656\,B   &              & $2.2052\pm0.1509$  & $\,\,\,2.242\pm0.124$    & $-9.063\pm0.131$   & $0.300$ & $17.9742\pm0.0037$ &                              &\\
	\hline
4660\,A   & 435259601    &  $3.8669\pm0.0198$  & $-2.978\pm0.024\,$   & $-27.508\pm0.018\,\,\,$  & $0.000$ & $14.0977\pm0.0029$ &                              &\\
4660\,B   &	             &  $3.6936\pm0.0848$  & $-3.177\pm0.096\,$   & $-27.031\pm0.078\,\,\,$  & $0.664$ & $14.1363\pm0.0028$ & $1.6649_{-0.0868}^{+0.1333}$ &\\
	\hline
4661\,A   & 441135051    & $3.6713\pm0.0429$  & $\,44.084\pm0.044$   & $-29.203\pm0.041\,\,\,$  & $0.294$ & $12.8209\pm0.0032$ &                              &\\
4661\,B	  &              & $3.5748\pm0.0333$  & $\,43.829\pm0.035$   & $-29.113\pm0.034\,\,\,$  & $0.174$ & $13.0830\pm0.0028$ & $0.0000_{-0.0000}^{+0.9791}$ &\\
	\hline
4668\,A   & 142938659    &  $9.4339\pm0.0237$  & $\,35.719\pm0.021$   & $-4.416\pm0.028$   & $0.036$ & $15.1633\pm0.0029$ &                              &\\
4668\,B   &	             &  $9.4221\pm0.0256$  & $\,35.240\pm0.023$   & $-4.313\pm0.032$   & $0.109$ & $15.1874\pm0.0028$ & $0.0357_{-0.0357}^{+0.0853}$ &\\
	\hline
4725\,A   & 26587613     &  $2.7530\pm0.0186$  & $-5.927\pm0.023\,$   & $-6.782\pm0.019$   & $0.097$ & $12.7149\pm0.0028$ & $0.1546_{-0.1344}^{+0.1427}$ &\\
4725\,B   &	             &  $2.8154\pm0.0183$  & $-5.882\pm0.020\,$   & $-6.747\pm0.017$   & $0.089$ & $12.7365\pm0.0029$ &                              &\\
4725\,C   &	             &  $2.6933\pm0.4404$  & $-5.353\pm0.507\,$   & $-7.231\pm0.470$   & $0.000$ & $19.7832\pm0.0060$ &                              &\\
	\hline
4735\,A   & 280292434    & $5.0151\pm0.0222$  & $-26.192\pm0.023\,\,\,$  & $-11.679\pm0.021\,\,\,$  & $0.131$ & $14.0006\pm0.0028$ & $1.3196_{-0.1858}^{+0.2093}$ &\\
4735\,B   &              & $5.0403\pm0.0293$  & $-26.315\pm0.032\,\,\,$  & $-12.548\pm0.028\,\,\,$  & $0.188$ & $14.6276\pm0.0029$ &                              &\\
	\hline
4776\,A   & 196286578     & $2.6684\pm0.0137$  & $\,20.123\pm0.012$   & $-25.989\pm0.012\,\,\,$  & $0.090$ & $12.0200\pm0.0028$ & $0.1915_{-0.1692}^{+0.1639}$ &\\
4776\,B   &               & $3.0587\pm0.4433$  & $\,19.911\pm0.209$   & $-25.010\pm0.207\,\,\,$  & $1.053$ & $16.7312\pm0.0049$ &                              &\\
	\hline
4781\,A   & 176242778     & $2.0566\pm0.0200$  & $-5.598\pm0.020\,$   & $-3.433\pm0.017$   & $0.178$ & $12.0275\pm0.0028$ &                              &\\
4781\,B   &               & $1.8676\pm0.0299$  & $-5.055\pm0.030\,$   & $-3.458\pm0.024$   & $0.252$ & $12.5995\pm0.0031$ & $1.4697_{-0.5250}^{+0.9211}$ &\\
	\hline
4800\,A   & 280307604     & $2.7972\pm0.0595$  & $-4.608\pm0.061\,$   & $-0.875\pm0.050$   & $0.358$ & $11.3544\pm0.0028$ &                              &\\
4800\,B   &	              & $2.4874\pm0.0403$  & $-4.454\pm0.043\,$   & $-0.196\pm0.034$   & $0.242$ & $12.4439\pm0.0028$ & $0.2717_{-0.1270}^{+0.1543}$ &\\
	\hline
4858\,A   & 262499797     & $5.0381\pm0.0331$  & $\,54.190\pm0.051$   & $-92.530\pm0.037\,\,\,$  & $0.000$ & $16.0502\pm0.0029$ & $1.3313_{-0.4418}^{+0.2620}$ &\texttt{\maltese}\\
4858\,B   &	              & $4.4085\pm0.5066$  & $\,51.909\pm0.838$   & $-93.231\pm0.611\,\,\,$  & $2.174$ & $20.0915\pm0.0065$ &                              &\\
	\hline
4944\,A   & 242190195     & $2.0978\pm0.0190$  & $\,\,\,6.390\pm0.017$    & $-22.878\pm0.023\,\,\,$  & $0.000$ & $13.4861\pm0.0028$ & $0.2996_{-0.1891}^{+0.1343}$ &\\
4944\,B   &	              & $1.1986\pm0.2024$  & $\,\,\,6.235\pm0.190$    & $-21.967\pm0.289\,\,\,$  & $1.154$ & $16.7061\pm0.0037$ &                              &\\
	\hline
4980\,A   & 200435203     & $1.5847\pm0.2332$  & $-2.801\pm0.260\,$   & $\,\,\,\,1.209\pm0.285$    & $2.650$ & $13.4775\pm0.0031$ & $0.1345_{-0.1345}^{+0.2571}$ &\\
4980\,B   &	              &  $1.2358\pm0.2898$  & $-2.762\pm0.327\,$   & $\,\,\,\,1.160\pm0.337$    & $0.939$ & $19.6680\pm0.0046$ &                              &\\
	\hline
5049\,A   & 261020738     & $2.0831\pm0.0148$  & $-5.973\pm0.008\,$   & $-7.287\pm0.014$   & $0.000$ & $13.4514\pm0.0028$ &                              &\\
5049\,B   &	              & $1.5070\pm0.1882$  & $-6.131\pm0.096\,$   & $-6.784\pm0.181$   & $0.820$ & $17.5603\pm0.0055$ & $1.1681_{-0.8362}^{+1.1354}$ &\\
	\hline
5053\,A   & 324426685     & $4.6027\pm0.0119$  & $-10.696\pm0.011\,\,\,$  & $\,\,\,\,7.541\pm0.014$    & $0.039$ & $13.4350\pm0.0028$ & $0.5288_{-0.1030}^{+0.1089}$ &\\
5053\,B   &	              & $4.7140\pm0.4026$  & $-9.869\pm0.381$         & $\,\,\,\,7.359\pm0.420$    & $1.289$ & $20.0350\pm0.0051$ &                              &\\
	\hline
5069\,A   & 381360757     & $4.4150\pm0.0179$  & $\,\,\,7.266\pm0.019$    & $-42.685\pm0.015\,\,\,$  & $0.074$ & $10.0745\pm0.0028$ &                              &\\
5069\,B   &	              & $4.2829\pm0.0559$  & $\,\,\,7.409\pm0.058$    & $-42.676\pm0.050\,\,\,$  & $0.000$ & $16.2016\pm0.0029$ & $0.6145_{-0.0369}^{+0.0771}$ &\\
	\hline
5076\,A   & 303432813     & $12.0845\pm0.0145\,\,\,$ & $168.778\pm0.016\,\,\,$  & $-176.012\pm0.013\,\,\,\,\,\,$ & $0.063$ & $10.5928\pm0.0028$ &                              &\\
 5076\,B  &	              & $11.9869\pm0.0749\,\,\,$ & $168.200\pm0.077\,\,\,$  & $-175.411\pm0.065\,\,\,\,\,\,$ & $0.080$ & $16.6446\pm0.0030$ & $0.0595_{-0.0165}^{+0.0126}$ &\\
	\hline
5099\,A   & 91481801      & $10.8711\pm0.0352\,\,\,$ & $\,52.215\pm0.040$   & $-24.639\pm0.043\,\,\,$  & $0.137$ & $\,\,\,7.2569\pm0.0028$  &                              &\\
5099\,B   &	              & $10.8657\pm0.1281\,\,\,$ & $\,54.508\pm0.154$   & $-22.473\pm0.158\,\,\,$  & $0.581$ & $14.2516\pm0.0043$ & $0.1870_{-0.1258}^{+0.0708}$ &\texttt{SHC2}\\
\hline
\end{tabular}}
\end{table*}
	
\setcounter{table}{2}
	
\begin{table*}[h]
\caption{continued}
\resizebox{\hsize}{!}{\begin{tabular}{ccccccccc}
\hline
TOI & TIC               & $\pi$ & $\mu_\alpha\,cos(\delta)$ & $\mu_\delta$ & $epsi$ & $G$ & $A_{\rm G}$ &\\
	&                   & [mas] & [mas/yr] & [mas/yr] & [mas] & [mag] & [mag] &\\
	\hline
5115\,A & 366499151     & $4.8429\pm0.0148$  & $97.723\pm0.016$   & $-115.749\pm0.012\,\,\,\,\,\,$ & $0.070$ & $12.6509\pm0.0028$ & $0.2213_{-0.1293}^{+0.1380}$ &\\
5115\,B &               & $5.1713\pm0.2641$  & $98.186\pm0.296$   & $-116.518\pm0.215\,\,\,\,\,\,$ & $0.475$ & $18.9321\pm0.0107$ &                              &\\
	\hline
5122\,A & 116668723     & $5.5538\pm0.0205$  & $21.011\pm0.019$   & $-28.224\pm0.011\,\,\,$  & $0.087$ & $10.2255\pm0.0028$ & $0.3897_{-0.2216}^{+0.2228}$ &\\
5122\,B &               & $3.3330\pm0.2519$  & $26.586\pm0.243$   & $-31.423\pm0.143\,\,\,$  & $1.271$ & $14.8256\pm0.0031$ &                              &\\
	\hline
5128\,A & 406495245     & $5.1833\pm0.0328$  & $-17.336\pm0.037\,\,\,\,$  & $-14.671\pm0.027\,\,\,$  & $0.166$ & $\,\,\,8.4712\pm0.0028$  & $0.1927_{-0.1724}^{+0.1601}$ &\texttt{SHC2}\\
5128\,B &               & $4.9147\pm0.0458$  & $-18.252\pm0.052\,\,\,\,$  & $-14.385\pm0.031\,\,\,$  & $0.224$ & $11.0733\pm0.0030$ &                              &\\
5128\,C &               & $5.1797\pm0.1001$  & $-18.281\pm0.117\,\,\,\,$  & $-14.065\pm0.087\,\,\,$  & $0.375$ & $17.0132\pm0.0029$ &                              &\\
	\hline
5129\,A & 101520163     & $4.9568\pm0.0195$  & $-12.410\pm0.026\,\,\,\,$  & $\,\,\,\,6.692\pm0.020$    & $0.127$ & $10.5628\pm0.0028$ &                              &\\
5129\,B &               & $4.9621\pm0.1061$  & $-12.933\pm0.132\,\,\,\,$  & $\,\,\,\,7.733\pm0.113$    & $0.106$ & $17.5724\pm0.0032$ & $0.0448_{-0.0199}^{+0.0183}$ &\\
	\hline
5130\,A & 75589027      & $9.9609\pm0.0218$        & $56.520\pm0.020$   & $-83.881\pm0.012\,\,\,$  & $0.089$ & $\,\,\,9.0289\pm0.0028$  & $0.1731_{-0.1731}^{+0.1663}$ &\\
5130\,B &               & $10.1243\pm0.1440\,\,\,$ & $55.451\pm0.146$   & $-83.841\pm0.081\,\,\,$  & $0.437$ & $16.9184\pm0.0040$ &                              &\\
	\hline
5148\,A & 291517604     & $4.1152\pm0.0166$  & $\,\,\,9.614\pm0.012$    & $\,\,\,\,8.079\pm0.012$    & $0.085$ & $10.5197\pm0.0028$ & $0.3045_{-0.2210}^{+0.2459}$ &\\
5148\,B &               & $3.8753\pm0.2469$  & $\,\,\,9.729\pm0.277$    & $\,\,\,\,8.798\pm0.248$    & $0.823$ & $18.6780\pm0.0060$ &                              &\\
	\hline
5176\,A & 437054764     & $3.7873\pm0.0369$  & $-71.336\pm0.042\,\,\,\,$  & $-27.146\pm0.033\,\,\,$  & $0.000$ & $15.7132\pm0.0028$ & $0.1735_{-0.0516}^{+0.0360}$ &\\
5176\,B &               & $4.4470\pm0.4272$  & $-72.973\pm0.511\,\,\,\,$  & $-26.743\pm0.294\,\,\,$  & $1.170$ & $19.1415\pm0.0040$ &                              &\\
	\hline
5181\,A & 346667887     & $2.1064\pm0.0139$  & $\,\,\,8.348\pm0.009$    & $\,\,\,\,2.698\pm0.012$    & $0.106$ & $12.2091\pm0.0028$ & $0.5724_{-0.2059}^{+0.1838}$ &\\
5181\,B &               & $2.3584\pm0.1212$  & $\,\,\,8.378\pm0.153$    & $\,\,\,\,3.626\pm0.121$    & $0.620$ & $17.2895\pm0.0034$ &                              &\\
	\hline
5233\,A & 259100469     & $4.0102\pm0.0122$  & $21.392\pm0.015$   & $\,87.246\pm0.015$   & $0.097$ & $11.6274\pm0.0028$ &                              &\\
5233\,B &               & $4.0602\pm0.0224$  & $21.487\pm0.029$   & $\,86.729\pm0.026$   & $0.000$ & $15.3101\pm0.0028$ & $0.0765_{-0.0379}^{+0.0616}$ &\\
	\hline
5241\,A & 305948428     & $2.3206\pm0.0109$  & $-11.861\pm0.007\,\,\,\,$  & $-12.065\pm0.011\,\,\,$  & $0.030$ & $12.6129\pm0.0028$ &                              &\\
5241\,B &               & $2.3021\pm0.0827$  & $-11.787\pm0.059\,\,\,\,$  & $-11.742\pm0.085\,\,\,$  & $0.000$ & $17.7455\pm0.0029$ & $0.2793_{-0.0998}^{+0.0611}$ &\texttt{SHC2}\\
	\hline
5242\,A & 426122503     & $2.9992\pm0.0106$  & $\,\,\,7.698\pm0.010$    & $-41.388\pm0.011\,\,\,$  & $0.000$ & $13.3451\pm0.0028$ &                              &\\
5242\,B &               & $2.9167\pm0.0889$  & $\,\,\,7.909\pm0.093$    & $-41.625\pm0.093\,\,\,$  & $0.395$ & $17.7388\pm0.0031$ & $1.2542_{-0.1871}^{+0.5444}$ &\\
	\hline
5273\,A & 172871230     & $2.2753\pm0.0105$  & $-8.755\pm0.012\,$   & $-17.781\pm0.012\,\,\,$  & $0.025$ & $12.7128\pm0.0028$ &                              &\\
5273\,B &               & $2.2845\pm0.0980$  & $-8.746\pm0.112\,$   & $-17.900\pm0.116\,\,\,$  & $0.503$ & $17.6242\pm0.0036$ & $1.6924_{-0.7669}^{+0.2772}$ &\\
	\hline
5285\,B & 250330564     & $5.3355\pm0.0191$  & $-52.315\pm0.019\,\,\,\,$  & $-82.913\pm0.018\,\,\,$  & $0.000$ & $13.7481\pm0.0028$ &                              &\\
5285\,A &               & $5.5292\pm0.2424$  & $-49.666\pm0.234\,\,\,\,$  & $-80.126\pm0.223\,\,\,$  & $1.765$ & $12.3317\pm0.0029$ & $1.0634_{-0.1319}^{+0.2244}$ &\\
	\hline
5287\,A & 909930	    & $2.5977\pm0.0274$  & $-8.536\pm0.029\,$   & $-29.633\pm0.023\,\,\,$  & $0.160$ & $11.1279\pm0.0028$ &                              &\\
5287\,B &               & $2.5694\pm0.0334$  & $-7.905\pm0.044\,$   & $-29.682\pm0.029\,\,\,$  & $0.170$ & $12.6902\pm0.0029$ & $0.0860_{-0.0860}^{+0.1468}$ &\texttt{SHC2}\\
	\hline
5291\,A & 250983039     & $4.6466\pm0.0969$  & $-4.909\pm0.103\,$   & $\,19.145\pm0.085$   & $0.665$ & $11.3714\pm0.0028$ &                              &\\
5291\,B &               & $4.0770\pm0.0586$  & $-3.875\pm0.063\,$   & $\,19.634\pm0.052$   & $0.393$ & $13.1258\pm0.0029$ & $0.5832_{-0.1195}^{+0.1463}$ &\\
	\hline
5292\,A & 33397739      & $2.7901\pm0.0416$  & $-8.733\pm0.041\,$   & $-6.862\pm0.039$   & $0.053$ & $15.8931\pm0.0029$ &                              &\\
5292\,B &               & $2.7656\pm0.1777$  & $-9.300\pm0.171\,$   & $-7.163\pm0.169$   & $0.308$ & $18.4652\pm0.0036$ & $0.0893_{-0.0151}^{+0.0516}$ &\texttt{SHC2}\\
	\hline
5293\,A & 250111245     & $6.1645\pm0.0282$  & $-17.131\pm0.031\,\,\,\,$  & $\,\,\,\,0.487\pm0.024$    & $0.000$ & $14.9622\pm0.0028$ &                              &\\
5293\,B &               & $6.2742\pm0.1330$  & $-16.497\pm0.152\,\,\,\,$  & $       -0.079\pm0.119$   & $0.000$ & $17.8662\pm0.0034$ & $0.2981_{-0.2981}^{+0.0374}$ &\\
	\hline
5294\,A & 422486295     & $3.0820\pm0.0250$  & $12.360\pm0.034$   & $-11.291\pm0.022\,\,\,$  & $0.171$ & $12.7674\pm0.0029$ & $0.1243_{-0.0625}^{+0.0625}$ &\texttt{\maltese}\\
5294\,B &               & $3.2861\pm0.0274$  & $12.513\pm0.031$   & $-10.580\pm0.023\,\,\,$  & $0.141$ & $12.9076\pm0.0028$ &                              &\\
	\hline
5296\,A & 241102583     & $3.7277\pm0.0155$  & $44.367\pm0.016$   & $-7.514\pm0.017$   & $0.019$ & $13.0509\pm0.0028$ &                              &\\
5296\,B &               & $3.7129\pm0.0181$  & $44.527\pm0.017$   & $-8.366\pm0.017$   & $0.000$ & $13.3199\pm0.0029$ & $0.6772_{-0.1489}^{+0.1384}$ &\\
	\hline
5319\,A & 246965431     & $16.4500\pm0.0169\,\,\,$ & $42.116\pm0.021$   & $-86.609\pm0.017\,\,\,$  & $0.035$ & $13.1190\pm0.0029$ & $0.7196_{-0.2637}^{+0.0785}$ &\\
5319\,B &               & $16.2534\pm0.0480\,\,\,$ & $41.583\pm0.058$   & $-84.732\pm0.042\,\,\,$  & $0.252$ & $14.5828\pm0.0029$ &                              &\\
	\hline
5324\,A & 408643099     & $3.5559\pm0.0185$  & $\,\,\,6.031\pm0.021$    & $-16.594\pm0.018\,\,\,$  & $0.000$ & $13.8681\pm0.0028$ &                              &\\
5324\,B &               & $3.5616\pm0.1574$  & $\,\,\,6.482\pm0.178$    & $-16.680\pm0.161\,\,\,$  & $0.000$ & $18.2873\pm0.0034$ & $0.1844_{-0.0676}^{+0.0856}$ &\texttt{SHC2}\\
	\hline
5335\,A & 381360750     & $3.0155\pm0.0168$  & $\,\,\,1.682\pm0.019$    & $-1.920\pm0.015$   & $0.000$ & $13.0734\pm0.0028$ &                              &\\
5335\,B &               & $3.0346\pm0.0900$  & $\,\,\,1.720\pm0.093$    & $-1.675\pm0.081$   & $0.000$ & $17.0253\pm0.0029$ & $0.4154_{-0.0258}^{+0.0389}$ &\\
	\hline
\end{tabular}}
\end{table*}
	
\setcounter{table}{2}
	
\begin{table*}[h]
\caption{continued}
\resizebox{\hsize}{!}{\begin{tabular}{ccccccccc}
\hline
TOI & TIC               & $\pi$ & $\mu_\alpha\,cos(\delta)$ & $\mu_\delta$ & $epsi$ & $G$ & $A_{\rm G}$ &\\
	&                   & [mas] & [mas/yr] & [mas/yr] & [mas] & [mag] & [mag] &\\
	\hline
5337\,A & 27294830      & $3.2720\pm0.0573$  & $12.289\pm0.065$   & $-15.651\pm0.051\,\,\,$  & $0.387$ & $13.7008\pm0.0028$ &                              &\\
5337\,B &               & $3.3119\pm0.1361$  & $12.110\pm0.145$   & $-15.954\pm0.115\,\,\,$  & $0.472$ & $17.4804\pm0.0035$ & $1.2545_{-0.2650}^{+0.3091}$ &\texttt{SHC2}\\
	\hline
5347\,A & 21119973	    & $2.9037\pm0.0227$  & $\,\,\,5.952\pm0.026$    & $-1.415\pm0.023$   & $0.062$ & $14.1743\pm0.0028$ &                              &\\
5347\,B &               & $2.7994\pm0.0467$  & $\,\,\,5.662\pm0.053$    & $-1.598\pm0.051$   & $0.246$ & $15.0675\pm0.0030$ & $0.0005_{-0.0005}^{+0.1516}$ &\\
	\hline
5385\,A & 85266608	    & $2.0927\pm0.0209$  & $-9.210\pm0.019\,$   & $\,\,\,\,0.858\pm0.019$    & $0.128$ & $11.2328\pm0.0028$ &                              &\\
5385\,B &               & $2.3919\pm0.1228$  & $-8.648\pm0.120\,$   & $\,\,\,\,0.414\pm0.131$    & $0.228$ & $17.7185\pm0.0031$ & $0.1974_{-0.0700}^{+0.0532}$ &\\
	\hline
5389\,A & 39143128	    & $5.1298\pm0.0565$  & $-33.023\pm0.044\,\,\,\,$  & $-46.700\pm0.040\,\,\,$  & $0.098$ & $15.9254\pm0.0028$ & $0.0027_{-0.0027}^{+0.0911}$ &\\
5389\,B &               & $7.6227\pm1.2444$  & $-32.538\pm0.691\,\,\,\,$  & $-47.623\pm0.631\,\,\,$  & $0.000$ & $20.1916\pm0.0067$ &                              &\\
	\hline
5390\,A & 8853478	    & $5.6819\pm0.0254$  & $-19.729\pm0.023\,\,\,\,$  & $-22.215\pm0.018\,\,\,$  & $0.121$ & $11.4856\pm0.0029$ & $0.3285_{-0.1335}^{+0.1238}$ &\\
5390\,B &               & $5.7284\pm0.1045$  & $-19.446\pm0.098\,\,\,\,$  & $-21.749\pm0.076\,\,\,$  & $0.401$ & $16.5204\pm0.0035$ &                              &\\
	\hline
5392\,A & 198512478	    & $20.6143\pm0.0135\,\,\,$ & $-61.343\pm0.018\,\,\,\,$  & $-95.375\pm0.016\,\,\,$  & $0.099$ & $\,\,\,8.6232\pm0.0028$  &                       &\\
5392\,B &               & $20.6001\pm0.0120\,\,\,$ & $-63.160\pm0.016\,\,\,\,$  & $-92.453\pm0.015\,\,\,$  & $0.089$ & $11.0908\pm0.0028$ & $0.0760_{-0.0670}^{+0.1458}$ &\\
	\hline
5397\,A & 18019350	    & $7.5004\pm0.0140$  & $-74.225\pm0.014\,\,\,\,$  & $-3.295\pm0.018$   & $0.074$ & $10.5880\pm0.0028$ &                              &\\
5397\,B &               & $7.5385\pm0.0190$  & $-74.593\pm0.018\,\,\,\,$  & $-3.226\pm0.023$   & $0.127$ & $12.9008\pm0.0028$ & $0.0642_{-0.0640}^{+0.1915}$ &\\
	\hline
5462\,A & 79971247	    & $3.4434\pm0.0285$  & $13.507\pm0.029$   & $-15.233\pm0.023\,\,\,$  & $0.155$ & $11.2564\pm0.0028$ & $0.2253_{-0.1391}^{+0.1629}$ &\\
5462\,B &               & $3.2245\pm0.3854$  & $12.478\pm0.367$   & $-15.741\pm0.314\,\,\,$  & $0.915$ & $18.5907\pm0.0103$ &                              &\\
	\hline
5518\,A & 386620582	    & $2.3023\pm0.0235$  & $-6.248\pm0.022\,$   & $-15.393\pm0.017\,\,\,$  & $0.124$ & $11.6631\pm0.0028$ &                              &\\
5518\,B &               & $2.3105\pm0.0226$  & $-6.472\pm0.021\,$   & $-15.261\pm0.016\,\,\,$  & $0.102$ & $12.0640\pm0.0028$ & $0.0718_{-0.0718}^{+0.0953}$ &\texttt{SHC2}\\
	\hline
5530\,A & 244170332	    & $16.4848\pm0.0202\,\,\,$ & $241.138\pm0.021\,\,\,$  & $-192.878\pm0.016\,\,\,\,\,\,$ & $0.080$ & $12.3423\pm0.0028$ & $0.1405_{-0.0893}^{+0.0394}$ &\\
5530\,B &               & $16.4699\pm0.0448\,\,\,$ & $240.601\pm0.052\,\,\,$  & $-192.594\pm0.031\,\,\,\,\,\,$ & $0.000$ & $15.2443\pm0.0028$ &                              &\\
	\hline
5540\,A & 456305193	    & $4.2720\pm0.0376$  & $-8.540\pm0.022\,$   & $-26.991\pm0.033\,\,\,$  & $0.331$ & $10.6287\pm0.0028$ & $0.2336_{-0.2336}^{+0.2540}$ &\\
5540\,B &               & $4.3668\pm0.0637$  & $-6.892\pm0.038\,$   & $-28.406\pm0.053\,\,\,$  & $0.456$ & $15.1140\pm0.0030$ &                              &\\
	\hline
5544\,A & 68893269	    & $8.8661\pm0.0767$  & $71.459\pm0.083$   & $-70.534\pm0.062\,\,\,$  & $0.609$ & $10.9148\pm0.0028$ & $0.3246_{-0.1587}^{+0.1370}$ &\\
5544\,B &               & $8.6165\pm0.1255$  & $65.979\pm0.183$   & $-70.416\pm0.152\,\,\,$  & $0.497$ & $16.9423\pm0.0041$ &                              &\\
	\hline
5551\,A & 387974148	    & $9.9811\pm0.0152$        & $-26.584\pm0.018\,\,\,\,$  & $\,\,\,\,0.712\pm0.011$    & $0.000$ & $13.4380\pm0.0028$ & $0.1840_{-0.0786}^{+0.0503}$ &\\
5551\,B &               & $10.2526\pm0.1734\,\,\,$ & $-26.815\pm0.212\,\,\,\,$  & $\,\,\,\,0.115\pm0.134$    & $0.460$ & $18.3372\pm0.0034$ &                              &\\
	\hline
5567\,A & 456335739     & $2.2963\pm0.0168$  & $-10.549\pm0.016\,\,\,\,$  & $-15.827\pm0.014\,\,\,$  & $0.000$ & $13.8018\pm0.0028$ &                              &\\
5567\,B &               & $2.3169\pm0.0177$  & $-10.504\pm0.017\,\,\,\,$  & $-15.825\pm0.015\,\,\,$  & $0.000$ & $13.9431\pm0.0028$ & $0.1687_{-0.1276}^{+0.1449}$ &\\
	\hline
5578\,A & 154563411	    & $6.2937\pm0.0103$  & $-45.050\pm0.013\,\,\,\,$  & $-8.363\pm0.012$   & $0.019$ & $13.0972\pm0.0028$ & $0.0764_{-0.0764}^{+0.1628}$ &\\
5578\,B &               & $6.2417\pm0.1382$  & $-45.283\pm0.176\,\,\,\,$  & $-7.819\pm0.162$   & $0.000$ & $18.5988\pm0.0034$ &                              &\\
	\hline
5606\,B & 171599496	    & $3.1460\pm0.0183$  & $\,\,\,7.539\pm0.017$    & $\,10.261\pm0.017$   & $0.000$ & $14.0042\pm0.0028$ &                              &\\
5606\,A &               & $3.1284\pm0.0184$  & $\,\,\,8.918\pm0.016$    & $\,10.936\pm0.015$   & $0.000$ & $13.6769\pm0.0028$ & $0.2998_{-0.1385}^{+0.1518}$ &\\
	\hline
5620\,A & 165414210	    & $4.5223\pm0.0142$  & $24.434\pm0.014$   & $-37.973\pm0.015\,\,\,$  & $0.059$ & $12.5514\pm0.0028$ & $0.3581_{-0.1477}^{+0.1789}$ &\\
5620\,B &               & $4.2225\pm0.1231$  & $24.734\pm0.138$   & $-37.485\pm0.158\,\,\,$  & $0.352$ & $17.7859\pm0.0046$ &                              &\\
	\hline
5626\,A & 376645976	    & $8.1152\pm0.0193$  & $-65.474\pm0.016\,\,\,\,$  & $-31.153\pm0.015\,\,\,$  & $0.138$ & $10.2611\pm0.0028$ & $0.1223_{-0.1179}^{+0.1212}$ &\\
5626\,B &               & $7.9906\pm0.3347$  & $-65.165\pm0.325\,\,\,\,$  & $-31.320\pm0.252\,\,\,$  & $1.288$ & $19.3796\pm0.0040$ &                              &\\
	\hline
5628\,A & 135100529	    & $9.0682\pm0.0445$  & $-170.769\pm0.040\,\,\,\,\,\,\,$ & $-28.445\pm0.042\,\,\,$  & $0.024$ & $15.5706\pm0.0028$ & $0.0695_{-0.0356}^{+0.0232}$ &\\
5628\,B &               & $8.9716\pm0.0974$  & $-170.202\pm0.094\,\,\,\,\,\,\,$ & $-28.622\pm0.088\,\,\,$  & $0.000$ & $17.1961\pm0.0030$ &                              &\\
	\hline
5630\,A & 316416562	    & $7.4542\pm0.0097$  & $17.601\pm0.010$   & $-46.845\pm0.013\,\,\,$  & $0.061$ & $10.8306\pm0.0028$ & $0.1443_{-0.1164}^{+0.1281}$ &\\
5630\,B &               & $7.7131\pm0.1289$  & $17.686\pm0.123$   & $-46.811\pm0.167\,\,\,$  & $0.413$ & $18.5666\pm0.0033$ &                              &\\
	\hline
5634\,A & 119585136	    & $3.0895\pm0.0585$  & $-60.799\pm0.056\,\,\,\,$  & $-25.488\pm0.052\,\,\,$  & $0.000$ & $15.9849\pm0.0028$ &                              &\\
5634\,B &               & $2.8990\pm0.2339$  & $-60.802\pm0.213\,\,\,\,$  & $-25.625\pm0.207\,\,\,$  & $0.274$ & $18.3031\pm0.0034$ & $0.0658_{-0.0658}^{+0.0381}$ &\\
	\hline
5657\,A & 441736986	    & $3.4188\pm0.0141$  & $13.655\pm0.018$   & $-3.616\pm0.016$   & $0.000$ & $14.0714\pm0.0028$ &                              &\\
5657\,B &               & $3.3722\pm0.0667$  & $13.755\pm0.084$   & $-3.287\pm0.071$   & $0.523$ & $16.6299\pm0.0031$ & $1.1054_{-0.1675}^{+0.2620}$ &\\
	\hline
5661\,A & 349312122	    & $2.3953\pm0.0140$  & $-0.231\pm0.016\,$   & $-8.205\pm0.015$   & $0.032$ & $13.4637\pm0.0028$ &                              &\\
5661\,B &               & $2.2516\pm0.0962$  & $-0.578\pm0.111\,$   & $-8.288\pm0.099$   & $0.252$ & $17.5767\pm0.0048$ & $1.7849_{-0.7356}^{+0.1846}$ &\\
\hline
\end{tabular}}
\end{table*}

\setcounter{table}{2}

\begin{table*}[h]
\caption{continued}
\resizebox{\hsize}{!}{\begin{tabular}{cccccccc}
\hline
CTOI=TIC     & $\pi$ & $\mu_\alpha\,cos(\delta)$ & $\mu_\delta$ & $epsi$ & $G$ & $A_{\rm G}$ &\\
             & [mas] & [mas/yr] & [mas/yr] & [mas] & [mag] & [mag] &\\
\hline
29106627\,A  & $7.0234\pm0.0153$  & $-28.853\pm0.014\,\,\,\,$  & $-7.702\pm0.012$   & $0.067$ & $11.7592\pm0.0028$ & $0.0871_{-0.0871}^{+0.1251}$ &\\
29106627\,B  & $6.5615\pm0.6192$  & $-29.722\pm0.567\,\,\,\,$  & $-8.361\pm0.429$   & $1.068$ & $19.8962\pm0.0064$ &                              &\\
\hline
51099561\,A  & $16.4075\pm0.0640\,\,\,$ & $109.789\pm0.056\,\,\,$  & $-76.446\pm0.053\,\,\,$  & $0.523$ & $\,\,\,8.2180\pm0.0028$  &                              &\\
51099561\,B  & $17.0098\pm0.0305\,\,\,$ & $109.463\pm0.028\,\,\,$  & $-71.636\pm0.025\,\,\,$  & $0.226$ & $13.0244\pm0.0029$ & $0.0866_{-0.0389}^{+0.0670}$ &\\
\hline
73496987\,A  & $8.1130\pm0.0116$  & $68.141\pm0.009$   & $-46.748\pm0.010\,\,\,$  & $0.077$ & $12.1571\pm0.0028$ &                              &\\
73496987\,B  & $8.3217\pm0.0379$  & $66.663\pm0.028$   & $-47.595\pm0.033\,\,\,$  & $0.268$ & $15.4132\pm0.0028$ & $0.3337_{-0.0519}^{+0.0459}$ &\\
\hline
80435273\,A  & $4.0520\pm0.0222$  & $12.708\pm0.025$   & $-7.036\pm0.018$   & $0.128$ & $11.0797\pm0.0028$ & $0.0847_{-0.0847}^{+0.1778}$ &\\
80435273\,B  & $4.0918\pm0.4537$  & $12.149\pm0.600$   & $-6.406\pm0.451$   & $1.314$ & $19.6122\pm0.0047$ &                              &\\
\hline
123496379\,A & $3.9263\pm0.0111$  & $-13.112\pm0.014\,\,\,\,$  & $-31.179\pm0.012\,\,\,$  & $0.069$ & $11.9313\pm0.0028$ &                              &\\
123496379\,B & $3.9183\pm0.0794$  & $-13.270\pm0.128\,\,\,\,$  & $-30.782\pm0.101\,\,\,$  & $0.695$ & $16.1765\pm0.0031$ & $0.1616_{-0.1496}^{+0.1172}$ &\texttt{SHC2}\\
\hline
125489144\,A & $6.5278\pm0.0157$  & $12.826\pm0.011$   & $\,35.341\pm0.011$   & $0.103$ & $12.5269\pm0.0028$ & $0.5694_{-0.1279}^{+0.1248}$ &\\
125489144\,B & $6.3939\pm0.0867$  & $12.868\pm0.062$   & $\,36.723\pm0.081$   & $0.434$ & $16.8067\pm0.0029$ &                              &\\
\hline
130718055\,A & $13.0123\pm0.0238\,\,\,$ & $35.782\pm0.026$   & $-15.263\pm0.012\,\,\,$  & $0.106$ & $10.1216\pm0.0028$ &                              &\\
130718055\,B & $13.0071\pm0.0272\,\,\,$ & $34.887\pm0.031$   & $-15.089\pm0.017\,\,\,$  & $0.078$ & $14.4251\pm0.0029$ & $0.0506_{-0.0506}^{+0.0189}$ &\\
\hline
154927164\,A & $5.2602\pm0.0367$  & $-51.116\pm0.053\,\,\,\,$  & $-19.599\pm0.044\,\,\,$  & $0.139$ & $16.3528\pm0.0030$ & $0.3115_{-0.0279}^{+0.0489}$ &\\
154927164\,B & $5.4197\pm0.1136$  & $-51.390\pm0.174\,\,\,\,$  & $-19.556\pm0.134\,\,\,$  & $0.458$ & $18.4543\pm0.0033$ &                              &\\
\hline
178143624\,A & $5.3204\pm0.0232$  & $-9.289\pm0.026\,$   & $-18.950\pm0.021\,\,\,$  & $0.165$ & $11.5675\pm0.0028$ &                              &\\
178143624\,B & $5.2906\pm0.0319$  & $-8.991\pm0.035\,$   & $-18.721\pm0.030\,\,\,$  & $0.105$ & $14.7117\pm0.0028$ & $0.2521_{-0.0350}^{+0.0489}$ &\\
\hline
199660056\,A & $6.2238\pm0.0126$  & $-31.712\pm0.014\,\,\,\,$  & $-49.181\pm0.017\,\,\,$  & $0.080$ & $11.6897\pm0.0028$ & $0.0976_{-0.0976}^{+0.1202}$ &\\
199660056\,B & $6.1721\pm0.1345$  & $-31.926\pm0.142\,\,\,\,$  & $-48.341\pm0.175\,\,\,$  & $0.543$ & $18.0681\pm0.0040$ &                              &\\
\hline
237204346\,A & $3.4136\pm0.0120$  & $19.232\pm0.016$   & $\,30.312\pm0.015$   & $0.099$ & $11.0394\pm0.0028$ & $0.1474_{-0.1474}^{+0.2028}$ &\\
237204346\,B & $3.6087\pm0.1718$  & $19.353\pm0.247$   & $\,30.544\pm0.202$   & $0.801$ & $18.8487\pm0.0055$ &                              &\\
\hline
241249530\,A & $2.9570\pm0.0153$  & $\,\,\,7.620\pm0.018$    & $-15.717\pm0.015\,\,\,$  & $0.101$ & $11.5593\pm0.0028$ & $0.3559_{-0.1790}^{+0.2022}$ &\\
241249530\,B & $3.1728\pm0.0905$  & $\,\,\,7.766\pm0.107$    & $-15.799\pm0.092\,\,\,$  & $0.125$ & $17.4606\pm0.0040$ &                              &\\
\hline
246976997\,A & $11.8090\pm0.0538\,\,\,$ & $41.975\pm0.064$   & $-110.825\pm0.051\,\,\,\,\,\,$ & $0.420$ & $12.7758\pm0.0028$ &                              &\\
246976997\,B & $11.8951\pm0.0161\,\,\,$ & $41.200\pm0.021$   & $-104.588\pm0.017\,\,\,\,\,\,$ & $0.000$ & $13.5784\pm0.0028$ & $0.0886_{-0.0640}^{+0.0520}$ &\\
\hline
257554718\,A & $5.2141\pm0.0192$  & $-4.911\pm0.019\,$   & $\,40.835\pm0.017$   & $0.118$ & $11.0185\pm0.0028$ &                              &\\
257554718\,B & $3.5395\pm0.2891$  & $-2.990\pm0.290\,$   & $\,42.110\pm0.274$   & $2.254$ & $17.4789\pm0.0035$ & $0.2212_{-0.0318}^{+0.1106}$ &\\
\hline
199572211\,A & $14.4304\pm0.0365\,\,\,$ & $-11.933\pm0.044\,\,\,\,$  & $-103.185\pm0.048\,\,\,\,\,\,$ & $0.406$ & $14.5327\pm0.0028$ & $0.4388_{-0.0767}^{+0.0341}$ &\\
199572211\,B & $14.4517\pm0.0634\,\,\,$ & $       -8.691\pm0.079\,$  & $-101.813\pm0.086\,\,\,\,\,\,$ & $0.621$ & $16.0663\pm0.0032$ &                              &\\
\hline
287643871\,A & $4.7152\pm0.0209$  & $23.466\pm0.013$   & $-32.590\pm0.016\,\,\,$  & $0.149$ & $10.0678\pm0.0028$ & $0.4076_{-0.3951}^{+0.2319}$ &\\
287643871\,B & $5.0993\pm0.0908$  & $22.196\pm0.068$   & $-36.095\pm0.126\,\,\,$  & $0.353$ &                    &                              &\\
\hline
320261550\,A & $2.6690\pm0.0885$  & $15.015\pm0.066$   & $-50.948\pm0.070\,\,\,$  & $0.745$ & $11.6670\pm0.0028$ & $0.2429_{-0.1167}^{+0.1791}$ &\\
320261550\,B & $3.0603\pm0.2235$  & $14.643\pm0.179$   & $-49.992\pm0.190\,\,\,$  & $0.429$ & $18.8900\pm0.0035$ &                              &\\
\hline
333792947\,A & $6.7063\pm0.0181$  & $-51.995\pm0.017\,\,\,\,$  & $-50.514\pm0.013\,\,\,$  & $0.098$ & $10.6233\pm0.0028$ & $0.0673_{-0.0673}^{+0.1337}$ &\\
333792947\,B & $6.8644\pm0.6154$  & $-51.334\pm0.842\,\,\,\,$  & $-51.407\pm0.489\,\,\,$  & $1.546$ & $20.2015\pm0.0064$ &                              &\\
\hline
336892053\,A & $5.0249\pm0.0218$  & $26.890\pm0.026$   & $-12.772\pm0.019\,\,\,$  & $0.137$ & $11.4605\pm0.0028$ & $0.0862_{-0.0862}^{+0.1418}$ &\\
336892053\,B & $4.7220\pm0.4689$  & $27.186\pm0.556$   & $-13.132\pm0.379\,\,\,$  & $1.909$ & $19.5777\pm0.0046$ &                              &\\
\hline
355640518\,A & $4.3590\pm0.0383$  & $\,\,\,7.527\pm0.050$    & $-41.367\pm0.041\,\,\,$  & $0.088$ & $15.5950\pm0.0029$ &                              &\\
355640518\,B & $4.1693\pm0.0912$  & $\,\,\,9.297\pm0.117$    & $-41.767\pm0.101\,\,\,$  & $0.295$ & $16.9860\pm0.0029$ & $0.2089_{-0.0233}^{+0.0287}$ &\\
\hline
359046756\,A & $6.4011\pm0.0238$  & $ -9.236\pm0.018\,$        & $-23.787\pm0.022\,\,\,$  & $0.202$ & $11.7817\pm0.0028$ & $0.2898_{-0.1102}^{+0.1307}$ &\\
359046756\,B & $6.2385\pm0.2553$  & $-10.245\pm0.180\,\,\,\,$  & $-24.988\pm0.208\,\,\,$  & $0.920$ & $18.9581\pm0.0074$ &                              &\\
\hline
376457352\,B & $8.4273\pm0.0258$  & $42.314\pm0.028$   & $\,39.584\pm0.022$   & $0.150$ & $10.4927\pm0.0028$ &                              &\\
376457352\,A & $8.4022\pm0.0156$  & $42.280\pm0.019$   & $\,39.894\pm0.014$   & $0.053$ & $\,\,\,9.2576\pm0.0028$  & $0.0044_{-0.0044}^{+0.1874}$ &\\
\hline
376973804\,A & $12.9222\pm0.0766\,\,\,$ & $23.591\pm0.091$   & $147.484\pm0.091$  & $0.876$ & $10.2497\pm0.0028$ & $0.4544_{-0.1821}^{+0.1616}$ &\\
376973804\,B & $12.3639\pm0.0160\,\,\,$ & $34.787\pm0.018$   & $152.890\pm0.020$  & $0.143$ & $11.9360\pm0.0028$ &                              &\\
\hline
379376771\,A & $3.4568\pm0.0149$  & $\,\,\,7.305\pm0.013$    & $-2.240\pm0.016$   & $0.000$ & $14.1406\pm0.0030$ & $0.4784_{-0.0430}^{+0.2691}$ &\texttt{SHC2}\\
379376771\,B & $2.2507\pm0.6848$  & $\,\,\,7.890\pm0.742$    & $-3.441\pm0.846$   & $1.793$ & $20.5039\pm0.0072$ &                              &\\
\hline
\end{tabular}}
\end{table*}

\setcounter{table}{2}

\begin{table*}[h]
\caption{continued}
\resizebox{\hsize}{!}{\begin{tabular}{cccccccc}
\hline
CTOI=TIC     & $\pi$ & $\mu_\alpha\,cos(\delta)$ & $\mu_\delta$ & $epsi$ & $G$ & $A_{\rm G}$ &\\
             & [mas] & [mas/yr] & [mas/yr] & [mas] & [mag] & [mag] &\\
\hline
388076422\,B & $8.5120\pm0.0163$  & $-2.430\pm0.020\,$   & $-5.109\pm0.021$   & $0.000$ & $14.2959\pm0.0028$ &                              &\\
388076422\,A & $8.5280\pm0.0135$  & $-2.238\pm0.016\,$   & $-5.556\pm0.017$   & $0.096$ & $\,\,\,9.5818\pm0.0028$  & $0.0827_{-0.0827}^{+0.1591}$ &\\
\hline
389041242\,A & $4.6607\pm0.0235$  & $-20.289\pm0.022\,\,\,\,$  & $-8.591\pm0.013$   & $0.081$ & $\,\,\,9.4602\pm0.0028$  & $0.3500_{-0.2975}^{+0.3226}$ &\\
389041242\,B & $3.7849\pm0.1270$  & $-21.288\pm0.122\,\,\,\,$  & $-8.247\pm0.071$   & $0.619$ & $15.3608\pm0.0030$ &                              &\\
\hline
415732733\,A & $2.5769\pm0.0179$  & $\,\,\,9.245\pm0.017$    & $-12.775\pm0.018\,\,\,$  & $0.148$ & $11.2274\pm0.0028$ &                              &\\
415732733\,B & $2.5354\pm0.0172$  & $\,\,\,9.723\pm0.016$    & $-12.393\pm0.016\,\,\,$  & $0.130$ & $11.6734\pm0.0028$ & $0.1052_{-0.1052}^{+0.1107}$ &\texttt{SHC2}\\
\hline
446044800\,A & $3.8032\pm0.0225$  & $10.967\pm0.031$   & $-6.842\pm0.021$   & $0.132$ & $13.0133\pm0.0028$ & $2.9435_{-0.1107}^{+0.4187}$ &\texttt{SHC2}\\
446044800\,B & $4.0823\pm0.5194$  & $11.482\pm0.721$   & $-7.463\pm0.478$   & $2.895$ & $19.3764\pm0.0122$ &                              &\\
\hline
\end{tabular}}
\end{table*} 
\setcounter{table}{3}

\begin{table*}
\caption{In this table we list for each detected companion (sorted by its identifier) the angular separation $\rho$ and position angle $PA$ to the associated (C)TOI, the difference between its parallax and that of the (C)TOI $\Delta \pi$ with its significance (in brackets calculated also by taking into account the Gaia astrometric excess noise), the differential proper motion $\mu_{\rm rel}$ of the companion relative to the (C)TOI with its significance, as well as its $cpm$-$index$. The last column indicates if the detected companion is not listed in the WDS ($\star$) and WDSS ($\star\star$) as companion(-candidate) of the (C)TOI.} \label{TAB_COMP_RELASTRO}
\resizebox{\hsize}{!}{\begin{tabular}{ccccccccc}
\hline
TOI   & $\rho$    & $PA$        & $\Delta\pi$ & $sig$-      & $\mu_{\rm rel}$ & $sig$-          & $cpm$-   & not in\\
      & [arcsec]  & [$^\circ$]  & [mas]       & $\Delta\pi$ & [mas/yr]        & $\mu_{\rm rel}$ & $index$  & WDS(S)\\
\hline
4580\,B        & $28.83923\pm0.00008\,\,\,$  & $147.04551\pm0.00015$ & $0.01\pm0.08$ & 0.1 (0.0) & $2.13\pm0.10$ & 21.4   & 257  &        \\
4601\,B        & $13.50476\pm0.00002\,\,\,$  & $\,\,\,37.30170\pm0.00010$  & $0.00\pm0.03$ & 0.1 (0.0) & $0.23\pm0.03$ & 7.4    & 133  & $\star$\\
4609\,B        & $22.37831\pm0.00003\,\,\,$  & $322.91646\pm0.00008$ & $0.07\pm0.04$ & 1.5 (0.3) & $0.45\pm0.05$ & 9.2    & 27   & $\star$\\
4642\,A        & $110.76534\pm0.00003\,\,\,\,\,\,$ & $238.42220\pm0.00002$ & $0.09\pm0.04$ & 2.4 (0.4) & $2.60\pm0.04$ & 68.3   & 117  &        \\
4656\,B        & $1.82101\pm0.00011$   & $324.92595\pm0.00355$ & $0.10\pm0.15$ & 0.6 (0.3) & $0.60\pm0.13$ & 4.6    & 30   & $\star$\\
4660\,B        & $5.68704\pm0.00007$   & $345.72785\pm0.00076$ & $0.17\pm0.09$ & 2.0 (0.3) & $0.52\pm0.08$ & 6.2    & 106  & $\star$\\
4661\,B        & $0.97469\pm0.00004$   & $\,\,\,67.26731\pm0.00251$  & $0.10\pm0.05$ & 1.8 (0.3) & $0.27\pm0.06$ & 4.8    & 390  & $\star$\\
4668\,B        & $20.91123\pm0.00003\,\,\,$  & $194.94794\pm0.00007$ & $0.01\pm0.03$ & 0.3 (0.1) & $0.49\pm0.03$ & 15.4   & 146  & $\star$\\
4725\,B        & $3.47521\pm0.00002$   & $333.51156\pm0.00038$ & $0.06\pm0.03$ & 2.4 (0.5) & $0.06\pm0.03$ & 2.0    & 315  & $\star$\\
4725\,C        & $20.21887\pm0.00036\,\,\,$  & $285.89303\pm0.00096$ & $0.06\pm0.44$ & 0.1 (0.1) & $0.73\pm0.49$ & 1.5    & 25   & $\star\star$\\
4735\,B        & $1.83179\pm0.00003$   & $352.79459\pm0.00099$ & $0.03\pm0.04$ & 0.7 (0.1) & $0.88\pm0.04$ & 25.0   & 66   & $\star$\\
4776\,B        & $1.22640\pm0.00026$   & $105.69781\pm0.01939$ & $0.39\pm0.44$ & 0.9 (0.3) & $1.00\pm0.21$ & 4.8    & 65   & $\star$\\
4781\,B        & $1.28545\pm0.00002$   & $358.63232\pm0.00124$ & $0.19\pm0.04$ & 5.3 (0.6) & $0.54\pm0.04$ & 15.1   & 23   & $\star$\\
4800\,B        & $6.13520\pm0.00005$   & $\,\,\,\,\,\,9.71246\pm0.00055$   & $0.31\pm0.07$ & 4.3 (0.7) & $0.70\pm0.06$ & 11.4   & 13   &        \\
4858\,B        & $36.33844\pm0.00051\,\,\,$  & $\,\,\,35.89644\pm0.00079$  & $0.63\pm0.51$ & 1.2 (0.3) & $2.39\pm0.82$ & 2.9    & 90   & $\star$\\
4944\,B        & $2.24130\pm0.00017$   & $\,\,\,75.89413\pm0.00438$  & $0.90\pm0.20$ & 4.4 (0.8) & $0.92\pm0.29$ & 3.2    & 50   & $\star$\\
4980\,B        & $14.86605\pm0.00035\,\,\,$  & $\,\,\,41.35600\pm0.00133$  & $0.35\pm0.37$ & 0.9 (0.1) & $0.06\pm0.43$ & 0.1    & 97   & $\star\star$\\
5049\,B        & $1.68305\pm0.00014$   & $214.33608\pm0.00390$ & $0.58\pm0.19$ & 3.1 (0.7) & $0.53\pm0.18$ & 3.0    & 35   & $\star\star$\\
5053\,B        & $7.72716\pm0.00036$   & $196.67239\pm0.00225$ & $0.11\pm0.40$ & 0.3 (0.1) & $0.85\pm0.38$ & 2.2    & 30   & $\star$\\
5069\,B        & $26.27029\pm0.00005\,\,\,$  & $258.30703\pm0.00010$ & $0.13\pm0.06$ & 2.3 (1.4) & $0.14\pm0.06$ & 2.3    & 604  & $\star$\\
5076\,B        & $26.00300\pm0.00006\,\,\,$  & $330.04099\pm0.00014$ & $0.10\pm0.08$ & 1.3 (0.8) & $0.83\pm0.07$ & 11.5   & 584  &        \\
5099\,B        & $3.17353\pm0.00013$   & $\,\,\,56.36330\pm0.00242$  & $0.01\pm0.13$ & 0.0 (0.0) & $3.15\pm0.16$ & 19.6   & 37   & $\star$\\
5115\,B        & $3.40570\pm0.00023$   & $\,\,\,93.42056\pm0.00264$  & $0.33\pm0.26$ & 1.2 (0.6) & $0.90\pm0.24$ & 3.7    & 339  & $\star$\\
5122\,B        & $3.64290\pm0.00018$   & $\,\,\,50.55511\pm0.00265$  & $2.22\pm0.25$ & 8.8 (1.7) & $6.43\pm0.22$ & 28.8   & 12   & $\star\star$\\
5128\,B        & $1.46947\pm0.00004$   & $356.95198\pm0.00211$ & $0.27\pm0.06$ & 4.8 (0.9) & $0.96\pm0.06$ & 15.4   & 48   &        \\
5128\,C        & $29.43633\pm0.00009\,\,\,$  & $\,\,\,31.95409\pm0.00018$  & $0.00\pm0.11$ & 0.0 (0.0) & $1.12\pm0.11$ & 9.8    & 41   & $\star$\\
5129\,B        & $12.54948\pm0.00010\,\,\,$  & $282.23270\pm0.00045$ & $0.01\pm0.11$ & 0.0 (0.0) & $1.16\pm0.12$ & 9.8    & 25   & $\star$\\
5130\,B        & $5.89943\pm0.00008$   & $343.33835\pm0.00112$ & $0.16\pm0.15$ & 1.1 (0.3) & $1.07\pm0.15$ & 7.3    & 189  & $\star$\\
5148\,B        & $4.81438\pm0.00017$   & $290.75719\pm0.00182$ & $0.24\pm0.25$ & 1.0 (0.3) & $0.73\pm0.25$ & 2.9    & 35   & $\star$\\
5176\,B        & $1.96225\pm0.00024$   & $328.45046\pm0.00857$ & $0.66\pm0.43$ & 1.5 (0.5) & $1.69\pm0.50$ & 3.4    & 91   & $\star$\\
5181\,B        & $1.65364\pm0.00010$   & $172.47774\pm0.00386$ & $0.25\pm0.12$ & 2.1 (0.4) & $0.93\pm0.12$ & 7.6    & 19   & $\star$\\
5233\,B        & $10.82636\pm0.00002\,\,\,$  & $\,\,\,24.42815\pm0.00013$  & $0.05\pm0.03$ & 2.0 (0.5) & $0.53\pm0.03$ & 17.5   & 341  &        \\
5241\,B        & $7.24523\pm0.00005$   & $230.81927\pm0.00045$ & $0.02\pm0.08$ & 0.2 (0.2) & $0.33\pm0.08$ & 3.9    & 101  & $\star$\\
5242\,B        & $3.49198\pm0.00007$   & $\,\,\,89.09553\pm0.00123$  & $0.08\pm0.09$ & 0.9 (0.2) & $0.32\pm0.09$ & 3.4    & 266  & $\star$\\
5273\,B        & $3.46950\pm0.00009$   & $197.39491\pm0.00144$ & $0.01\pm0.10$ & 0.1 (0.0) & $0.12\pm0.12$ & 1.0    & 333  & $\star$\\
5285\,A        & $5.80929\pm0.00019$   & $\,\,\,42.12495\pm0.00192$  & $0.19\pm0.24$ & 0.8 (0.1) & $3.85\pm0.23$ & 16.8   & 50   & $\star$\\
5287\,B        & $1.58830\pm0.00004$   & $\,\,\,12.51592\pm0.00173$  & $0.03\pm0.04$ & 0.7 (0.1) & $0.63\pm0.05$ & 12.0   & 97   & $\star$\\
5291\,B        & $6.44139\pm0.00009$   & $359.71348\pm0.00091$ & $0.57\pm0.11$ & 5.0 (0.7) & $1.14\pm0.12$ & 9.8    & 35   &        \\
5292\,B        & $9.42981\pm0.00016$   & $\,\,\,51.57081\pm0.00095$  & $0.02\pm0.18$ & 0.1 (0.1) & $0.64\pm0.18$ & 3.7    & 36   & $\star$\\
5293\,B        & $3.57016\pm0.00012$   & $\,\,\,79.26403\pm0.00164$  & $0.11\pm0.14$ & 0.8 (0.8) & $0.85\pm0.14$ & 6.0    & 40   & $\star$\\
5294\,B        & $1.21682\pm0.00003$   & $308.05059\pm0.00141$ & $0.20\pm0.04$ & 5.5 (0.9) & $0.73\pm0.03$ & 22.3   & 46   & $\star$\\
5296\,B        & $4.92901\pm0.00002$   & $130.99644\pm0.00024$ & $0.01\pm0.02$ & 0.6 (0.5) & $0.87\pm0.02$ & 36.1   & 104  & $\star$\\
\hline
\end{tabular}}
\end{table*}

\setcounter{table}{3}

\begin{table*}
\caption{continued}
\resizebox{\hsize}{!}{\begin{tabular}{ccccccccc}
\hline
TOI   & $\rho$    & $PA$        & $\Delta\pi$ & $sig$-      & $\mu_{\rm rel}$ & $sig$-          & $cpm$-   & not in\\
      & [arcsec]  & [$^\circ$]  & [mas]       & $\Delta\pi$ & [mas/yr]        & $\mu_{\rm rel}$ & $index$  & WDS(S)\\\hline
5319\,B        & $3.03736\pm0.00004$   & $205.52996\pm0.00080$ & $0.20\pm0.05$ & 3.9 (0.8) & $1.95\pm0.05$ & 41.8   & 98   & $\star$\\
5324\,B        & $5.89184\pm0.00013$   & $175.53782\pm0.00145$ & $0.01\pm0.16$ & 0.0 (0.0) & $0.46\pm0.18$ & 2.6    & 77   & $\star$\\
5335\,B        & $5.76672\pm0.00007$   & $277.59977\pm0.00065$ & $0.02\pm0.09$ & 0.2 (0.2) & $0.25\pm0.08$ & 3.0    & 20   & $\star$\\
5337\,B        & $1.96436\pm0.00009$   & $180.78058\pm0.00369$ & $0.04\pm0.15$ & 0.3 (0.1) & $0.35\pm0.14$ & 2.6    & 113  & $\star$\\
5347\,B        & $1.99340\pm0.00004$   & $120.99555\pm0.00122$ & $0.10\pm0.05$ & 2.0 (0.4) & $0.34\pm0.06$ & 5.9    & 35   & $\star$\\
5385\,B        & $6.69990\pm0.00012$   & $172.76240\pm0.00088$ & $0.30\pm0.12$ & 2.4 (1.0) & $0.72\pm0.13$ & 5.7    & 25   & $\star$\\
5389\,B        & $12.55341\pm0.00085\,\,\,$  & $335.79706\pm0.00254$ & $2.49\pm1.25$ & 2.0 (2.0) & $1.04\pm0.65$ & 1.6    & 110  & $\star$\\
5390\,B        & $2.53375\pm0.00007$   & $201.00858\pm0.00195$ & $0.05\pm0.11$ & 0.4 (0.1) & $0.55\pm0.08$ & 6.4    & 108  & $\star$\\
5392\,B        & $11.47138\pm0.00002\,\,\,$  & $155.49563\pm0.00010$ & $0.01\pm0.02$ & 0.8 (0.1) & $3.44\pm0.02$ & 152.6  & 65   &        \\
5397\,B        & $26.05021\pm0.00002\,\,\,$  & $\,\,\,92.51147\pm0.00005$  & $0.04\pm0.02$ & 1.6 (0.3) & $0.37\pm0.02$ & 16.2   & 398  &        \\
5462\,B        & $4.45576\pm0.00032$   & $293.35550\pm0.00410$ & $0.22\pm0.39$ & 0.6 (0.2) & $1.15\pm0.36$ & 3.2    & 35   & $\star$\\
5518\,B        & $2.58880\pm0.00002$   & $320.10931\pm0.00052$ & $0.01\pm0.03$ & 0.3 (0.1) & $0.26\pm0.03$ & 9.0    & 128  & $\star$\\
5530\,B        & $11.07450\pm0.00003\,\,\,$  & $\,\,\,26.67810\pm0.00021$  & $0.01\pm0.05$ & 0.3 (0.2) & $0.61\pm0.05$ & 11.6   & 1016 &        \\
5540\,B        & $2.80797\pm0.00005$   & $209.63715\pm0.00076$ & $0.09\pm0.07$ & 1.3 (0.2) & $2.17\pm0.05$ & 41.3   & 26   & $\star$\\
5544\,B        & $2.49444\pm0.00010$   & $178.70509\pm0.00322$ & $0.25\pm0.15$ & 1.7 (0.3) & $5.48\pm0.20$ & 27.3   & 36   & $\star$\\
5551\,B        & $37.16531\pm0.00015\,\,\,$  & $235.64238\pm0.00021$ & $0.27\pm0.17$ & 1.6 (0.6) & $0.64\pm0.15$ & 4.4    & 83   & $\star$\\
5567\,B        & $8.04728\pm0.00002$   & $\,\,\,13.06587\pm0.00013$  & $0.02\pm0.02$ & 0.8 (0.8) & $0.05\pm0.02$ & 1.9    & 844  & $\star$\\
5578\,B        & $29.88560\pm0.00012\,\,\,$  & $178.37760\pm0.00028$ & $0.05\pm0.14$ & 0.4 (0.4) & $0.59\pm0.16$ & 3.6    & 155  & $\star$\\
5606\,A        & $7.56543\pm0.00002$   & $316.50689\pm0.00016$ & $0.02\pm0.03$ & 0.7 (0.7) & $1.54\pm0.02$ & 66.1   & 17   & $\star\star$\\
5620\,B        & $3.32727\pm0.00012$   & $316.19626\pm0.00210$ & $0.30\pm0.12$ & 2.4 (0.8) & $0.57\pm0.15$ & 3.7    & 157  & $\star$\\
5626\,B        & $58.72491\pm0.00021\,\,\,$  & $150.48596\pm0.00025$ & $0.12\pm0.34$ & 0.4 (0.1) & $0.35\pm0.31$ & 1.1    & 412  & $\star$\\
5628\,B        & $22.65965\pm0.00008\,\,\,$  & $268.52158\pm0.00021$ & $0.10\pm0.11$ & 0.9 (0.9) & $0.59\pm0.10$ & 5.8    & 582  &        \\
5630\,B        & $23.09375\pm0.00010\,\,\,$  & $119.79307\pm0.00028$ & $0.26\pm0.13$ & 2.0 (0.6) & $0.09\pm0.13$ & 0.7    & 1093 & $\star$\\
5634\,B        & $3.80667\pm0.00022$   & $187.72980\pm0.00250$ & $0.19\pm0.24$ & 0.8 (0.5) & $0.14\pm0.21$ & 0.6    & 963  & $\star$\\
5657\,B        & $2.19933\pm0.00007$   & $162.58982\pm0.00189$ & $0.05\pm0.07$ & 0.7 (0.1) & $0.34\pm0.07$ & 4.6    & 82   & $\star$\\
5661\,B        & $2.82858\pm0.00008$   & $\,\,\,95.81690\pm0.00162$  & $0.14\pm0.10$ & 1.5 (0.5) & $0.36\pm0.11$ & 3.2    & 46   & $\star\star$\\
\hline
\end{tabular}}
\end{table*}

\setcounter{table}{3}

\begin{table*}
\caption{continued}
\resizebox{\hsize}{!}{\begin{tabular}{ccccccccc}
\hline
CTOI  & $\rho$    & $PA$        & $\Delta\pi$ & $sig$-      & $\mu_{\rm rel}$ & $sig$-          & $cpm$-   & not in\\
      & [arcsec]  & [$^\circ$]  & [mas]       & $\Delta\pi$ & [mas/yr]        & $\mu_{\rm rel}$ & $index$  & WDS(S)\\\hline
29106627\,B    & $10.37999\pm0.00043\,\,\,$  & $247.52702\pm0.00233$ & $0.46\pm0.62$ & 0.7 (0.4) & $1.09\pm0.52$ & 2.1    & 56   & $\star$\\
51099561\,B    & $28.08447\pm0.00006\,\,\,$  & $140.51172\pm0.00011$ & $0.60\pm0.07$ & 8.5 (1.0) & $4.82\pm0.06$ & 82.2   & 55   &        \\
73496987\,B    & $4.58833\pm0.00003$   & $236.88089\pm0.00034$ & $0.21\pm0.04$ & 5.3 (0.7) & $1.70\pm0.03$ & 55.4   & 97   & $\star$\\
80435273\,B    & $5.99073\pm0.00038$   & $248.53835\pm0.00308$ & $0.04\pm0.45$ & 0.1 (0.0) & $0.84\pm0.52$ & 1.6    & 34   & $\star$\\
123496379\,B   & $1.93228\pm0.00008$   & $221.65257\pm0.00233$ & $0.01\pm0.08$ & 0.1 (0.0) & $0.43\pm0.11$ & 4.0    & 158  & $\star$\\
125489144\,B   & $2.35644\pm0.00006$   & $302.51923\pm0.00144$ & $0.13\pm0.09$ & 1.5 (0.3) & $1.38\pm0.08$ & 16.9   & 55   & $\star$\\
130718055\,B   & $81.02682\pm0.00003\,\,\,$  & $\,\,\,53.97081\pm0.00002$  & $0.01\pm0.04$ & 0.1 (0.0) & $0.91\pm0.04$ & 22.8   & 84   & $\star$\\
154927164\,B   & $34.99119\pm0.00010\,\,\,$  & $176.98906\pm0.00021$ & $0.16\pm0.12$ & 1.3 (0.3) & $0.28\pm0.18$ & 1.5    & 396  &        \\
178143624\,B   & $5.51984\pm0.00003$   & $116.02747\pm0.00027$ & $0.03\pm0.04$ & 0.8 (0.1) & $0.38\pm0.04$ & 9.1    & 111  & $\star$\\
199572211\,B   & $1.62613\pm0.00007$   & $330.12105\pm0.00251$ & $0.02\pm0.07$ & 0.3 (0.0) & $3.52\pm0.09$ & 38.4   & 59   & $\star$\\
199660056\,B   & $5.78148\pm0.00012$   & $\,\,\,\,\,\,1.87768\pm0.00105$   & $0.05\pm0.14$ & 0.4 (0.1) & $0.87\pm0.17$ & 5.0    & 134  & $\star$\\
237204346\,B   & $5.77104\pm0.00017$   & $167.28659\pm0.00178$ & $0.20\pm0.17$ & 1.1 (0.2) & $0.26\pm0.21$ & 1.2    & 275  & $\star$\\
241249530\,B   & $4.93044\pm0.00008$   & $204.99356\pm0.00097$ & $0.22\pm0.09$ & 2.4 (1.2) & $0.17\pm0.11$ & 1.6    & 209  & $\star$\\
246976997\,B   & $15.02432\pm0.00005\,\,\,$  & $\,\,\,48.12895\pm0.00018$  & $0.09\pm0.06$ & 1.5 (0.2) & $6.28\pm0.05$ & 116.4  & 37   &        \\
257554718\,B   & $15.75378\pm0.00022\,\,\,$  & $\,\,\,\,\,\,1.02047\pm0.00079$   & $1.67\pm0.29$ & 5.8 (0.7) & $2.31\pm0.29$ & 8.1    & 36   & $\star$\\
287643871\,B   & $0.80477\pm0.00010$   & $\,\,\,57.02908\pm0.00956$  & $0.38\pm0.09$ & 4.1 (1.0) & $3.73\pm0.12$ & 30.6   & 22   &        \\
320261550\,B   & $12.08541\pm0.00020\,\,\,$  & $170.15320\pm0.00084$ & $0.39\pm0.24$ & 1.6 (0.4) & $1.03\pm0.20$ & 5.1    & 103  & $\star$\\
333792947\,B   & $19.17506\pm0.00048\,\,\,$  & $137.05849\pm0.00148$ & $0.16\pm0.62$ & 0.3 (0.1) & $1.11\pm0.64$ & 1.7    & 131  & $\star$\\
336892053\,B   & $18.23203\pm0.00036\,\,\,$  & $127.74226\pm0.00103$ & $0.30\pm0.47$ & 0.6 (0.2) & $0.47\pm0.46$ & 1.0    & 129  & $\star$\\
355640518\,B   & $2.55404\pm0.00008$   & $\,\,\,77.63236\pm0.00151$  & $0.19\pm0.10$ & 1.9 (0.6) & $1.81\pm0.13$ & 14.4   & 47   & $\star$\\
359046756\,B   & $6.08946\pm0.00017$   & $\,\,\,71.40152\pm0.00162$  & $0.16\pm0.26$ & 0.6 (0.2) & $1.57\pm0.20$ & 7.9    & 33   & $\star$\\
376457352\,A   & $80.99367\pm0.00002\,\,\,$  & $309.73802\pm0.00001$ & $0.03\pm0.03$ & 0.8 (0.2) & $0.31\pm0.03$ & 11.9   & 372  &        \\
376973804\,B   & $4.87687\pm0.00008$   & $\,\,\,62.90705\pm0.00090$  & $0.56\pm0.08$ & 7.1 (0.6) & $12.43\pm0.09$ & 133.9 & 25   &        \\
379376771\,B   & $11.53677\pm0.00053\,\,\,$  & $280.56036\pm0.00263$ & $1.21\pm0.68$ & 1.8 (0.6) & $1.34\pm0.83$ & 1.6    & 12   & $\star\star$\\
388076422\,A   & $40.07783\pm0.00002\,\,\,$  & $327.26457\pm0.00003$ & $0.02\pm0.02$ & 0.8 (0.2) & $0.49\pm0.03$ & 18.1   & 24   & $\star$\\
389041242\,B   & $6.35605\pm0.00009$   & $297.30798\pm0.00067$ & $0.88\pm0.13$ & 6.8 (1.4) & $1.06\pm0.12$ & 8.8    & 42   & $\star\star$\\
415732733\,B   & $2.00504\pm0.00002$   & $\,\,\,59.95759\pm0.00063$  & $0.04\pm0.02$ & 1.7 (0.2) & $0.61\pm0.02$ & 25.9   & 52   &        \\
446044800\,B   & $4.16024\pm0.00049$   & $\,\,\,80.62962\pm0.00380$  & $0.28\pm0.52$ & 0.5 (0.1) & $0.81\pm0.59$ & 1.4    & 33   & $\star\star$\\
\hline
\end{tabular}}
\end{table*} 
\setcounter{table}{4}

\begin{table*}
\caption{In this table we show the equatorial coordinates ($\alpha$, $\delta$ for epoch 2016.0) of all detected co-moving companions (sorted by their identifier) together with their derived absolute G-band magnitude $M_{\rm G}$, projected separation $sep$, mass, and effective temperature $T_{\rm eff}$. The flags for all companions, as defined in the text, are listed in the last column of this table.}\label{TAB_COMP_PROPS} \resizebox{\hsize}{!}{\begin{tabular}{ccccccccc}
\hline
TOI          & $\alpha$        & $\delta$        & $M_{\rm G}$             & $sep$  & $mass$                  &  $T_{\rm eff}$         & Flags \\
             & [$^\circ$]      & [$^\circ$]      & [mag]                   & [au]   & [$M_{\odot}$]           &  [K]                   & \\
\hline
4580\,B      & 242.42571706220 & 65.82126284363  & $13.01_{-0.33}^{+0.03}$ & $1951$ & $0.14_{-0.01}^{+0.01}$  & $2834_{-5}^{+109}$     & \texttt{BPRP}                             \\
4601\,B      & 75.81661135820  & 24.22625173722  & $7.14_{-1.05}^{+0.06}$  & $2968$ & $0.65_{-0.01}^{+0.02}$  & $4418_{-155}^{+19}$    & \texttt{SHC2} \texttt{BPRP} \texttt{AEN} \\
4609\,B      & 70.37018229541  & 21.26269129279  & $2.03_{-0.29}^{+0.29}$  & $7690$ & $1.65_{-0.25}^{+0.18}$  & $7335_{-861}^{+283}$   & \texttt{SHC2} \texttt{BPRP} \texttt{AEN} \\
4642\,A      & 30.85827552638  & 6.79974137729   & $10.88_{-0.02}^{+0.01}$ & $2610$ & $0.30_{-0.01}^{+0.01}$  & $3180_{-28}^{+29}$     &  \texttt{BPRP} \texttt{AEN}               \\
4656\,B      & 2.91065995420   & -38.00985393244 & $10.54_{-0.18}^{+0.15}$ & $791 $ & $0.30_{-0.11}^{+0.14}$  & $3378_{-286}^{+340}$   &                                           \\
4660\,B      & 51.99938517259  & 12.52494179753  & $5.52_{-0.14}^{+0.09}$  & $1471$ & $0.87_{-0.05}^{+0.07}$  & $5194_{-143}^{+177}$   &  \texttt{BPRP} \texttt{AEN} \texttt{NSS} \\
4661\,B      & 40.52044679463  & -17.50163406726 & $6.16_{-0.98}^{+0.03}$  & $265 $ & $0.73_{-0.18}^{+0.20}$  & $5050_{-1018}^{+892}$  &                                           \\
4668\,B      & 52.64552564136  & -35.20932766785 & $10.02_{-0.09}^{+0.04}$ & $2217$ & $0.30_{-0.03}^{+0.06}$  & $3052_{-12}^{+55}$     &  \texttt{BPRP} \texttt{AEN}               \\
4725\,B      & 108.00844677469 & 15.34977248291  & $4.77_{-0.15}^{+0.14}$  & $1262$ & $0.93_{-0.08}^{+0.07}$  & $5867_{-291}^{+264}$   &  \texttt{BPRP} \texttt{AEN}               \\
4725\,C      & 108.00329177814 & 15.35044647794  & $11.83_{-0.15}^{+0.14}$ & $7344$ & $0.20_{-0.01}^{+0.01}$  & $3135_{-17}^{+18}$     & \texttt{inter} \texttt{BPRP}              \\
4735\,B      & 100.89948849807 & -5.55143166603  & $9.50_{-0.21}^{+0.19}$  & $365 $ & $0.40_{-0.12}^{+0.18}$  & $3572_{-325}^{+570}$   &                                           \\
4776\,B      & 125.55481160934 & -25.06763051147 & $8.67_{-0.17}^{+0.17}$  & $460 $ & $0.52_{-0.02}^{+0.02}$  & $3703_{-48}^{+49}$     & \texttt{inter}                            \\
4781\,B      & 121.78212444434 & -18.72655477400 & $2.41_{-0.93}^{+0.53}$  & $625 $ & $1.43_{-0.35}^{+0.64}$  & $6521_{-1167}^{+3523}$ & \texttt{BPRP}                             \\
4800\,B      & 114.70887523979 & 4.35804999669   & $4.17_{-0.17}^{+0.14}$  & $2193$ & $0.98_{-0.09}^{+0.11}$  & $6102_{-258}^{+327}$   &  \texttt{BPRP} \texttt{AEN}               \\
4858\,B      & 109.89066528814 & -55.72025719269 & $12.27_{-0.27}^{+0.45}$ & $7213$ & $0.17_{-0.02}^{+0.01}$  & $3082_{-61}^{+33}$     & \texttt{inter} \texttt{BPRP}              \\
4944\,B      & 210.95261189649 & -43.27352661884 & $8.02_{-0.14}^{+0.20}$  & $1068$ & $0.59_{-0.03}^{+0.01}$  & $3932_{-99}^{+29}$     & \texttt{inter} \texttt{BPRP}              \\
4980\,B      & 70.41543544324  & -45.48176870066 & $10.53_{-0.42}^{+0.35}$ & $9381$ & $0.31_{-0.04}^{+0.05}$  & $3336_{-68}^{+82}$     & \texttt{inter} \texttt{BPRP}              \\
5049\,B      & 226.11142327044 & -66.20345005275 & $7.75_{-1.14}^{+0.84}$  & $808 $ & $0.58_{-0.06}^{+0.07}$  & $4133_{-229}^{+257}$   & \texttt{SHC2}                             \\
5053\,B      & 285.66558091709 & -77.58455457430 & $12.82_{-0.11}^{+0.11}$ & $1679$ & $0.15_{-0.01}^{+0.01}$  & $3007_{-16}^{+16}$     & \texttt{inter} \texttt{BPRP}              \\
5069\,B      & 46.91343005792  & 15.32319842199  & $8.72_{-0.08}^{+0.04}$  & $5950$ & $0.50_{-0.05}^{+0.05}$  & $3557_{-74}^{+69}$     &  \texttt{BPRP} \texttt{AEN}               \\
5076\,B      & 50.50666435934  & 17.24544286280  & $11.95_{-0.02}^{+0.02}$ & $2152$ & $0.22_{-0.02}^{+0.01}$  & $2923_{-4}^{+6}$       &  \texttt{BPRP}                            \\
5099\,B      & 44.79434887878  & 19.99021854274  & $10.23_{-0.08}^{+0.13}$ & $292 $ & $0.33_{-0.13}^{+0.16}$  & $3410_{-279}^{+351}$   &                                           \\
5115\,B      & 127.15546132112 & 10.24941505705  & $12.14_{-0.14}^{+0.13}$ & $703 $ & $0.18_{-0.01}^{+0.01}$  & $3098_{-16}^{+17}$     & \texttt{inter} \texttt{BPRP}              \\
5122\,B      & 85.68320979997  & 30.76355816456  & $8.16_{-0.23}^{+0.23}$  & $656 $ & $0.57_{-0.02}^{+0.03}$  & $3850_{-66}^{+101}$    & \texttt{inter} \texttt{BPRP}              \\
5128\,B      & 143.31761521004 & 9.16882639535   & $4.49_{-0.17}^{+0.18}$  & $284 $ & $1.01_{-0.14}^{+0.27}$  & $5588_{-2826}^{+976}$  & \texttt{SHC2}                             \\
5128\,C      & 143.32202070526 & 9.17535654113   & $10.66_{-0.17}^{+0.18}$ & $5679$ & $0.25_{-0.03}^{+0.12}$  & $3014_{-33}^{+52}$     & \texttt{BPRP} \texttt{AEN}                \\
5129\,B      & 99.69889042928  & 29.09006190937  & $11.07_{-0.03}^{+0.03}$ & $2532$ & $0.30_{-0.01}^{+0.02}$  & $3123_{-21}^{+5}$      & \texttt{BPRP} \texttt{AEN}                \\
5130\,B      & 85.51589366676  & 28.24274993609  & $11.74_{-0.17}^{+0.18}$ & $592 $ & $0.20_{-0.01}^{+0.01}$  & $3146_{-22}^{+20}$     & \texttt{inter} \texttt{BPRP}              \\
5148\,B      & 261.23900112910 & -58.63258006051 & $11.45_{-0.25}^{+0.23}$ & $1170$ & $0.22_{-0.01}^{+0.01}$  & $3181_{-28}^{+31}$     & \texttt{inter} \texttt{BPRP}              \\
5176\,B      & 135.01919020509 & 13.27416633022  & $11.86_{-0.05}^{+0.06}$ & $518 $ & $0.19_{-0.01}^{+0.01}$  & $3132_{-7}^{+6}$       & \texttt{inter}                            \\
5181\,B      & 280.67753974839 & 21.69936519062  & $9.73_{-0.19}^{+0.21}$  & $785 $ & $0.38_{-0.16}^{+0.18}$  & $3532_{-367}^{+527}$   &                                           \\
5233\,B      & 291.17630715999 & 69.12792534863  & $8.25_{-0.07}^{+0.04}$  & $2700$ & $0.60_{-0.01}^{+0.01}$  & $3813_{-23}^{+33}$     &  \texttt{BPRP}                            \\
5241\,B      & 295.48207248825 & 23.12611920517  & $9.30_{-0.07}^{+0.11}$  & $3122$ & $0.50_{-0.01}^{+0.04}$  & $3483_{-46}^{+58}$     & \texttt{SHC2} \texttt{BPRP} \texttt{AEN}  \\
5242\,B      & 287.65966543260 & 35.49570988510  & $8.86_{-0.55}^{+0.19}$  & $1164$ & $0.44_{-0.04}^{+0.02}$  & $3957_{-178}^{+301}$   & \texttt{BPRP} \texttt{AEN}                \\
5273\,B      & 299.87847460642 & 41.74115164172  & $7.76_{-0.28}^{+0.77}$  & $1525$ & $0.55_{-0.04}^{+0.05}$  & $4157_{-361}^{+291}$   & \texttt{BPRP}                             \\
5285\,A      & 340.97018194418 & -1.83736466652  & $4.85_{-0.23}^{+0.14}$  & $1089$ & $0.91_{-0.04}^{+0.06}$  & $5481_{-187}^{+341}$   &  \texttt{BPRP} \texttt{AEN}               \\
5287\,B      & 328.03878268310 & -8.41923684736  & $4.67_{-0.15}^{+0.11}$  & $611 $ & $0.94_{-0.12}^{+0.11}$  & $5831_{-2378}^{+780}$  & \texttt{SHC2}                             \\
5291\,B      & 338.73321754730 & -11.36144610900 & $5.59_{-0.16}^{+0.13}$  & $1386$ & $0.84_{-0.04}^{+0.05}$  & $5240_{-156}^{+229}$   &  \texttt{BPRP} \texttt{AEN}               \\
5292\,B      & 344.22443464007 & -8.11812010696  & $10.70_{-0.07}^{+0.04}$ & $3380$ & $0.34_{-0.01}^{+0.05}$  & $3187_{-1}^{+9}$       & \texttt{SHC2} \texttt{BPRP}               \\
5293\,B      & 355.82965463346 & -2.04490969867  & $11.43_{-0.04}^{+0.3}$  & $579 $ & $0.27_{-0.02}^{+0.01}$  & $3019_{-134}^{+5}$     & \texttt{BPRP} \texttt{AEN}                \\
5294\,B      & 351.46744366346 & 4.45510285997   & $6.03_{-0.07}^{+0.07}$  & $395 $ & $0.74_{-0.17}^{+0.21}$  & $5130_{-1026}^{+881}$  &                                           \\
\hline
\end{tabular}}
\end{table*}

\setcounter{table}{4}

\begin{table*}
\caption{continued}
\resizebox{\hsize}{!}{\begin{tabular}{ccccccccc}
\hline
TOI     & $\alpha$  & $\delta$  & $M_{\rm G}$ & $sep$ & $mass$       &  $T_{\rm eff}$ & Flags \\
& [$^\circ$] & [$^\circ$] & [mag]   & [au]  & [$M_{\odot}$] &  [K]       & \\
\hline
5296\,B      & 328.97935135277 & -0.93830596959  & $5.52_{-0.14}^{+0.15}$  & $1322$ & $0.84_{-0.05}^{+0.05}$  & $5369_{-209}^{+214}$   &  \texttt{BPRP} \texttt{AEN}               \\
5319\,B      & 35.21316612425  & 23.51966741399  & $10.89_{-0.08}^{+0.27}$ & $185 $ & $0.32_{-0.01}^{+0.01}$  & $3101_{-6}^{+4}$       &  \texttt{BPRP}                            \\
5324\,B      & 35.91130263336  & 15.48132143882  & $10.67_{-0.09}^{+0.07}$ & $1657$ & $0.35_{-0.01}^{+0.01}$  & $3210_{-44}^{+38}$     & \texttt{SHC2} \texttt{BPRP} \texttt{AEN}  \\
5335\,B      & 46.95072702962  & 15.35929469431  & $8.82_{-0.05}^{+0.03}$  & $1912$ & $0.55_{-0.05}^{+0.01}$  & $3570_{-22}^{+53}$     &  \texttt{BPRP} \texttt{AEN}               \\
5337\,B      & 77.55608612838  & 19.63953587476  & $10.82_{-0.32}^{+0.27}$ & $600 $ & $0.25_{-0.10}^{+0.14}$  & $3280_{-294}^{+288}$   &                                           \\
5347\,B      & 36.13723265414  & 24.10135004532  & $7.28_{-0.16}^{+0.17}$  & $687 $ & $0.70_{-0.05}^{+0.01}$  & $4179_{-201}^{+104}$   &  \texttt{BPRP}                            \\
5385\,B      & 160.63097788008 & 28.19677946157  & $9.26_{-0.06}^{+0.08}$  & $3202$ & $0.50_{-0.05}^{+0.02}$  & $3502_{-26}^{+58}$     & \texttt{BPRP} \texttt{AEN}                \\
5389\,B      & 168.79981491846 & 39.36209426242  & $13.74_{-0.10}^{+0.03}$ & $2447$ &                         &                        & \texttt{BPRP} \texttt{WD}                 \\
5390\,B      & 142.44400894727 & 30.06224781742  & $11.08_{-0.13}^{+0.14}$ & $446 $ & $0.25_{-0.10}^{+0.15}$  & $3241_{-311}^{+306}$   &                                           \\
5392\,B      & 212.62573741757 & 72.58684434661  & $7.57_{-0.15}^{+0.07}$  & $556 $ & $0.65_{-0.01}^{+0.01}$  & $4085_{-82}^{+136}$    &  \texttt{BPRP}                            \\
5397\,B      & 172.02349593924 & 39.67232940285  & $7.19_{-0.20}^{+0.07}$  & $3473$ & $0.70_{-0.01}^{+0.01}$  & $4271_{-107}^{+177}$   &  \texttt{BPRP} \texttt{AEN}               \\
5462\,B      & 91.09747991820  & 29.07376227197  & $11.05_{-0.17}^{+0.15}$ & $1294$ & $0.25_{-0.02}^{+0.02}$  & $3238_{-26}^{+30}$     & \texttt{inter} \texttt{BPRP}              \\
5518\,B      & 124.15244115120 & 16.39297090092  & $3.84_{-0.10}^{+0.08}$  & $1124$ & $1.07_{-0.17}^{+0.11}$  & $6306_{-162}^{+244}$   & \texttt{SHC2}  \texttt{BPRP} \texttt{AEN} \\
5530\,B      & 6.49077239528   & -5.69492423465  & $11.46_{-0.04}^{+0.09}$ & $672 $ & $0.21_{-0.01}^{+0.02}$  & $3109_{-16}^{+52}$     &  \texttt{BPRP}                            \\
5540\,B      & 117.73647336421 & 67.04252126585  & $8.03_{-0.26}^{+0.24}$  & $657 $ & $0.59_{-0.03}^{+0.03}$  & $3930_{-111}^{+54}$     & \texttt{inter} \texttt{BPRP}              \\
5544\,B      & 74.70084177919  & 18.16834758492  & $11.36_{-0.14}^{+0.16}$ & $281 $ & $0.22_{-0.01}^{+0.01}$  & $3192_{-19}^{+17}$     & \texttt{inter}                            \\
5551\,B      & 87.11664355630  & 18.17370176834  & $13.15_{-0.06}^{+0.08}$ & $3724$ & $0.13_{-0.01}^{+0.01}$  & $2959_{-12}^{+9}$      & \texttt{inter} \texttt{BPRP}              \\
5567\,B      & 135.14019040998 & 49.45367095110  & $5.65_{-0.15}^{+0.13}$  & $3504$ & $0.83_{-0.05}^{+0.03}$  & $5223_{-175}^{+212}$   &  \texttt{BPRP} \texttt{AEN}               \\
5578\,B      & 189.26659149982 & 75.91214058224  & $12.52_{-0.17}^{+0.08}$ & $4748$ & $0.16_{-0.01}^{+0.01}$  & $3050_{-12}^{+22}$     & \texttt{inter} \texttt{BPRP}              \\
5606\,A      & 156.89236585321 & 22.42776119652  & $5.88_{-0.16}^{+0.14}$  & $2405$ & $0.79_{-0.04}^{+0.02}$  & $5162_{-182}^{+219}$   &  \texttt{BPRP} \texttt{AEN}               \\
5620\,B      & 159.70788368198 & 27.41616709350  & $10.70_{-0.18}^{+0.15}$ & $736 $ & $0.29_{-0.02}^{+0.02}$  & $3303_{-29}^{+35}$     & \texttt{inter} \texttt{BPRP}              \\
5626\,B      & 198.30818830247 & 31.62420359821  & $13.80_{-0.13}^{+0.12}$ & $7236$ & $0.11_{-0.01}^{+0.01}$  & $2871_{-16}^{+17}$     & \texttt{inter} \texttt{BPRP}              \\
5628\,B      & 185.72128171212 & 18.83653832030  & $11.91_{-0.03}^{+0.04}$ & $2499$ &                         &                        & \texttt{BPRP} \texttt{WD}                 \\
5630\,B      & 236.46116377899 & 40.37166786062  & $12.78_{-0.13}^{+0.12}$ & $3098$ & $0.15_{-0.01}^{+0.01}$  & $3012_{-17}^{+19}$     & \texttt{inter} \texttt{BPRP} \texttt{AEN} \\
5634\,B      & 175.64347526468 & 20.96424081778  & $10.41_{-0.06}^{+0.08}$ & $1232$ & $0.38_{-0.05}^{+0.02}$  & $3254_{-27}^{+31}$     & \texttt{SHC2} \texttt{BPRP} \texttt{AEN}  \\
5657\,B      & 258.83784456604 & 76.46840498296  & $7.75_{-0.27}^{+0.17}$  & $643 $ & $0.55_{-0.04}^{+0.03}$  & $4281_{-166}^{+274}$   & \texttt{BPRP}                             \\
5661\,B      & 222.36941815490 & 20.61525504777  & $7.29_{-0.19}^{+0.74}$  & $1181$ & $0.60_{-0.05}^{+0.04}$  & $4431_{-485}^{+177}$   & \texttt{BPRP}                             \\
\hline
\end{tabular}}
\end{table*}

\setcounter{table}{4}

\begin{table*}
\caption{continued}
\resizebox{\hsize}{!}{\begin{tabular}{ccccccccc}
\hline
CTOI     & $\alpha$  & $\delta$  & $M_{\rm G}$ & $sep$ & $mass$       &  $T_{\rm eff}$ & Flags \\
& [$^\circ$] & [$^\circ$] & [mag]   & [au]  & [$M_{\odot}$] &  [K]       & \\
\hline
29106627\,B  & 118.32161546570 & 44.02391278176  & $14.04_{-0.13}^{+0.09}$ & $1478$ &                         &                        & \texttt{BPRP} \texttt{WD}                 \\
51099561\,B  & 108.78812382154 & 57.27062098263  & $9.13_{-0.07}^{+0.04}$  & $1712$ & $0.53_{-0.03}^{+0.01}$  & $3447_{-27}^{+59}$     &  \texttt{BPRP}                            \\
73496987\,B  & 154.72975675913 & -40.36148352393 & $9.66_{-0.05}^{+0.06}$  & $566 $ & $0.48_{-0.14}^{+0.01}$  & $3290_{-60}^{+14}$     &  \texttt{BPRP} \texttt{AEN}               \\
80435273\,B  & 5.15744205473   & 28.00127695429  & $12.57_{-0.18}^{+0.09}$ & $1478$ & $0.16_{-0.01}^{+0.01}$  & $3043_{-13}^{+25}$     & \texttt{inter} \texttt{BPRP}              \\
123496379\,B & 282.71383958329 & 45.83846316355  & $10.06_{-0.12}^{+0.15}$ & $492 $ & $0.35_{-0.14}^{+0.15}$  & $3441_{-288}^{+367}$   &                                           \\
125489144\,B & 354.93318323256 & 40.59673766213  & $10.31_{-0.13}^{+0.13}$ & $361 $ & $0.34_{-0.02}^{+0.02}$  & $3379_{-25}^{+25}$     & \texttt{inter} \texttt{BPRP}              \\
130718055\,B & 203.49089405471 & -2.22223237509  & $9.98_{-0.02}^{+0.06}$  & $6227$ & $0.40_{-0.01}^{+0.01}$  & $3330_{-52}^{+11}$     &  \texttt{BPRP}                            \\
154927164\,B & 205.60279245879 & 87.58146256826  & $11.75_{-0.06}^{+0.04}$ & $6652$ & $0.20_{-0.01}^{+0.01}$  & $3145_{-5}^{+7}$       & \texttt{inter} \texttt{BPRP} \texttt{AEN} \\
178143624\,B & 50.13941241190  & 39.34963790918  & $8.12_{-0.05}^{+0.04}$  & $1037$ & $0.55_{-0.01}^{+0.01}$  & $3922_{-24}^{+29}$     &  \texttt{BPRP}                            \\
199572211\,B & 241.51892772090 & 60.94859587341  & $11.95_{-0.04}^{+0.08}$ & $113 $ & $0.18_{-0.06}^{+0.09}$  & $3057_{-295}^{+251}$   &                                           \\
199660056\,B & 251.04498228102 & 56.37254080691  & $11.94_{-0.13}^{+0.10}$ & $929 $ & $0.19_{-0.01}^{+0.01}$  & $3122_{-12}^{+16}$     & \texttt{inter} \texttt{BPRP}              \\
237204346\,B & 293.32503811565 & 76.92013397731  & $11.37_{-0.21}^{+0.15}$ & $1691$ & $0.22_{-0.01}^{+0.01}$  & $3191_{-18}^{+28}$     & \texttt{inter} \texttt{BPRP}              \\
241249530\,B & 95.30793309240  & 53.29377686598  & $9.46_{-0.21}^{+0.18}$  & $1667$ & $0.42_{-0.02}^{+0.02}$  & $3520_{-43}^{+48}$     & \texttt{inter} \texttt{BPRP}              \\
246976997\,B & 36.02698104174  & 20.69574023571  & $8.86_{-0.06}^{+0.07}$  & $1272$ & $0.50_{-0.01}^{+0.01}$  & $3646_{-48}^{+31}$     &  \texttt{BPRP} \texttt{AEN}               \\
257554718\,B & 46.32519910846  & -21.99770274178 & $10.48_{-0.12}^{+0.04}$ & $3021$ & $0.40_{-0.03}^{+0.05}$  & $3174_{-11}^{+71}$     & \texttt{BPRP}                             \\
287643871\,B & 326.83165923610 & 34.85106117909  &                         & $171 $ &                         &                        & \texttt{noGmag}                           \\
320261550\,B & 295.26073495396 & -54.45474892921 & $10.78_{-0.20}^{+0.14}$ & $4528$ &                         &                        & \texttt{BPRP} \texttt{WD}                 \\
333792947\,B & 120.18676748700 & 36.78096025094  & $14.27_{-0.14}^{+0.07}$ & $2859$ &                         &                        & \texttt{BPRP} \texttt{WD}                 \\
336892053\,B & 17.97484538835  & 2.62290831567   & $13.00_{-0.15}^{+0.09}$ & $3628$ & $0.14_{-0.01}^{+0.01}$  & $2981_{-13}^{+22}$     & \texttt{inter} \texttt{BPRP}              \\
355640518\,B & 79.93016535003  & 48.13604250820  & $10.11_{-0.04}^{+0.04}$ & $586 $ & $0.40_{-0.01}^{+0.03}$  & $3272_{-9}^{+9}$       &  \texttt{BPRP}                            \\
359046756\,B & 182.72407801785 & 35.75623726018  & $12.70_{-0.14}^{+0.12}$ & $951 $ & $0.15_{-0.01}^{+0.01}$  & $3024_{-17}^{+20}$     & \texttt{inter} \texttt{BPRP}              \\
376457352\,A & 8.28485614305   & 7.58589364789   & $3.87_{-0.19}^{+0.01}$  & $9611$ & $1.02_{-0.10}^{+0.14}$  & $6120_{-326}^{+383}$   &  \texttt{BPRP} \texttt{AEN}               \\
376973804\,B & 284.06813293803 & 55.60692088124  & $7.04_{-0.17}^{+0.19}$  & $377 $ & $0.69_{-0.02}^{+0.01}$  & $4293_{-69}^{+118}$    & \texttt{inter} \texttt{BPRP} \texttt{AEN} \\
379376771\,B & 30.26309237626  & 68.99613421738  & $12.72_{-0.27}^{+0.05}$ & $3337$ & $0.15_{-0.01}^{+0.01}$  & $3021_{-7}^{+39}$      & \texttt{inter} \texttt{BPRP}              \\
388076422\,A & 317.88377986622 & 68.41144007368  & $4.14_{-0.16}^{+0.09}$  & $4708$ & $1.05_{-0.10}^{+0.10}$  & $6132_{-333}^{+347}$   &  \texttt{BPRP} \texttt{AEN}               \\
389041242\,B & 142.01983463221 & 5.81979539158   & $8.35_{-0.33}^{+0.30}$  & $1364$ & $0.55_{-0.03}^{+0.04}$  & $3796_{-87}^{+136}$    & \texttt{inter} \texttt{BPRP}              \\
415732733\,B & 142.03379819381 & 78.94502728362  & $3.63_{-0.12}^{+0.11}$  & $778 $ & $1.10_{-0.15}^{+0.15}$  & $6393_{-226}^{+256}$   & \texttt{SHC2} \texttt{BPRP}               \\
446044800\,B & 71.46590406097  & 31.83394502732  & $9.33_{-0.42}^{+0.12}$  & $1094$ & $0.44_{-0.01}^{+0.05}$  & $3551_{-28}^{+87}$     & \texttt{inter} \texttt{BPRP}              \\
\hline
\end{tabular}}
\end{table*} 

\end{document}